\documentclass[10pt,journal,compsoc]{IEEEtran}

\ifCLASSOPTIONcompsoc
  \usepackage[nocompress]{cite}
\else
  \usepackage{cite}
\fi

\ifCLASSINFOpdf
  \usepackage[pdftex]{graphicx}
\else
\fi
\newcommand{\etal}{\textit{et al.}}
\usepackage{amsmath}
\usepackage{amssymb}
\usepackage{multirow}
\usepackage{subfigure}
\usepackage{color}
\usepackage{xcolor}
\usepackage{makecell}
\usepackage{setspace}
\usepackage{booktabs} 

\newcommand{\SYT}[1]{\textcolor{black}{{}#1}}
\newcommand{\SYTnew}[1]{\textcolor{black}{{}#1}}
\newcommand{\sytneww}[1]{\textcolor{black}{{}#1}}
\newcommand{\sytminor}[1]{\textcolor{black}{{}#1}}

\newcommand{\hbmajor}[1]{{\color{black}            {#1}}}
\newcommand{\hbminor}[1]{{\color{black}            {#1}}}

\begin{document}

\title{Human Motion Transfer with 3D Constraints and Detail Enhancement}

\author{Yang-Tian Sun, Qian-Cheng Fu, Yue-Ren Jiang, Zitao Liu, Yu-Kun Lai, Hongbo Fu, and Lin~Gao\IEEEauthorrefmark{1}
\thanks{\IEEEauthorrefmark{1} Corresponding Author is Lin Gao (gaolin@ict.ac.cn).}
\IEEEcompsocitemizethanks{
\IEEEcompsocthanksitem Y.-T. Sun, Y.-R. Jiang and L. Gao are with the Beijing Key Laboratory of Mobile Computing and Pervasive Device, Institute of Computing Technology, Chinese Academy of Sciences, Beijing, China, and also with the University of Chinese Academy of Sciences, Beijing, China.\protect\\
E-mail:\{sunyangtian, gaolin\}@ict.ac.cn, jiangyueren15@mails.ucas.ac.cn

\IEEEcompsocthanksitem Q.-C. Fu is with the Department of Computer Science, Boston University
\protect\\
E-mail:qcfu@bu.edu

\IEEEcompsocthanksitem Z. Liu is with the TAL Education Group, Beijing, China and also with the Guangdong Institute of Smart Education, Jinan University, Guangzhou, China.
\protect\\
E-mail:liuzitao@tal.com 

\IEEEcompsocthanksitem Y.-K. Lai is with the Visual Computing Group, School of Computer Science and Informatics, Cardiff University, Wales, UK.
\protect\\
E-mail:LaiY4@cardiff.ac.uk

\IEEEcompsocthanksitem H. Fu is with the School of Creative Media, City University of Hong Kong.
\protect\\
E-mail: hongbofu@cityu.edu.hk 

}
}

\markboth{IEEE Transactions on Analysis and Machine Intelligence,~Vol.~xx, No.~xx, April~2021}%
{Lin \MakeLowercase{\textit{et al.}}: Human Motion Transfer with 3D Constraints and Detail Enhancement}

\IEEEtitleabstractindextext{%
\begin{abstract}
We propose a new method for realistic human motion transfer using a generative adversarial network (GAN), which generates a motion video of a target character imitating actions of a source character, while maintaining high authenticity of the generated results. We tackle the problem by decoupling and recombining the posture information and appearance information of both the source and target characters. The innovation of our approach lies in the use of
the projection of a reconstructed 3D human model as the condition of GAN to better maintain the structural integrity of transfer results in different poses. We further introduce a detail enhancement net to enhance the details of transfer results by exploiting the details in real source frames. Extensive experiments show that our approach yields better results both qualitatively and quantitatively than the state-of-the-art methods.
\end{abstract}

}

\maketitle

\begin{figure*}[t]
\begin{center}
\includegraphics[width=0.95\textwidth]{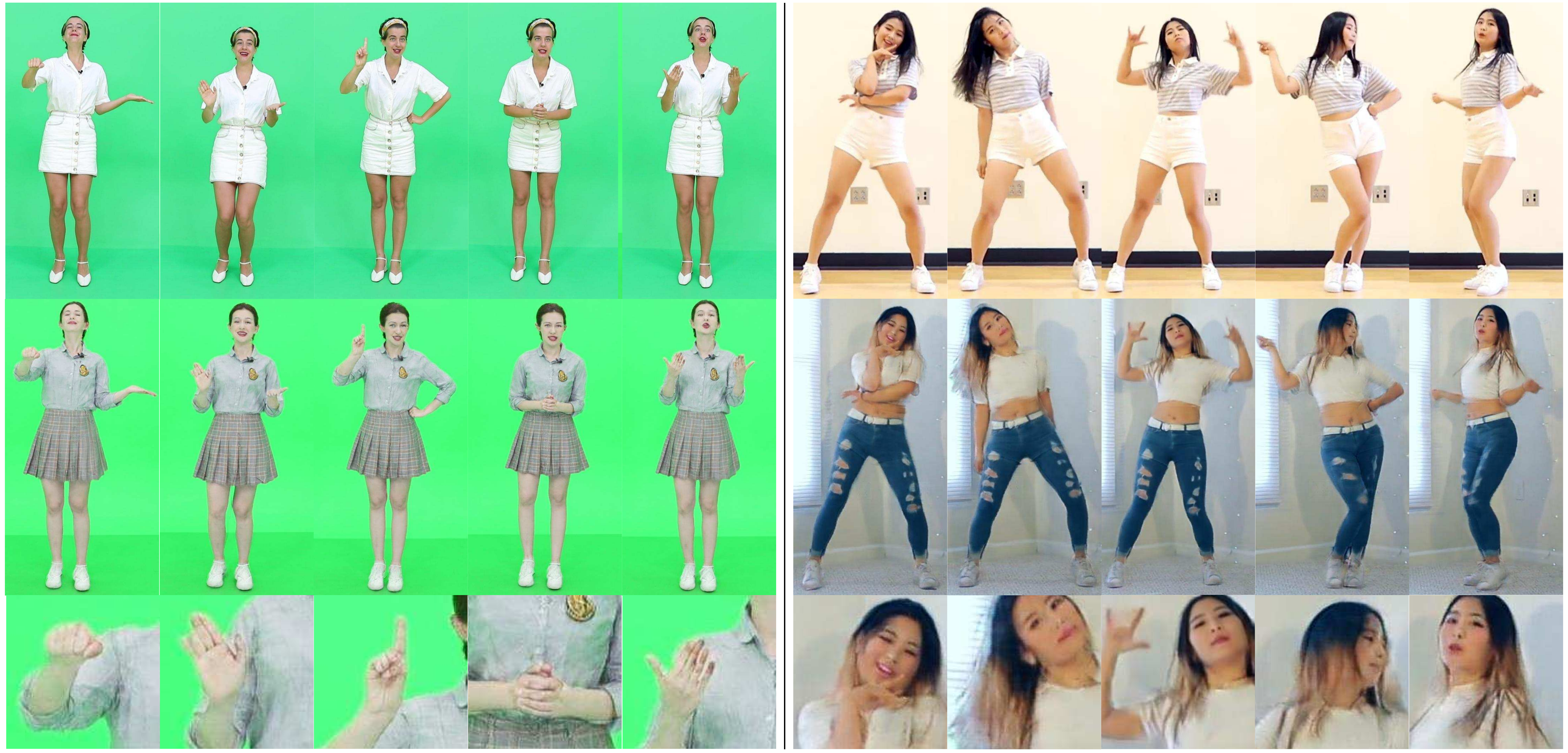}
\end{center}
\vspace{-3mm}
\caption{Given two monocular video clips, our method is able to transfer the motion of a source character (Top) to a target character (Middle), with realistic details (Bottom). \sytneww{We zoom in the enhanced details for better visualization.}
}
\vspace{-3mm}
\label{fig:teaser}
\end{figure*}

\IEEEdisplaynontitleabstractindextext

%
\IEEEpeerreviewmaketitle

\IEEEraisesectionheading{\section{Introduction}\label{sec:introduction}}

\IEEEPARstart{V}{ideo}-based
 human motion transfer is an interesting but challenging research problem.
Given two monocular video clips, one for a source subject and the other for a target subject, the goal of this problem
is to synthesize a video by transferring 
the motion from the source person to the target, while maintaining the target person's appearance. Specifically,
in the synthesized video, the subject should have
the same motion as the source person, and the same appearance as the target person (including human clothes and background).
Note this problem might also be seen as an appearance transfer problem if viewed from a single frame perspective. However, since from a holistic perspective, the goal is to transfer the motion from the source image domain to the target, we define this task as motion transfer. To achieve this, it is essential to produce high-quality image-to-image translation of individual frames, while ensuring temporal coherence.

The difficulty of this problem is how to effectively decouple and recombine the posture information and appearance information of the source and target characters. Based on generative adversarial networks (GANs), a powerful tool for high-quality image-to-image translation, Chan \etal~\cite{chan18:every_dance_now} proposed to first learn a mapping from a 2D pose to a subject image from a 
target video, and then use the pose of a 
source subject as the input to the learned mapping for video synthesis. However, due to the difference between the source and target poses, this approach often results in noticeable artifacts, especially for the self-occlusion of body parts.

Observing that the self-occlusion issue is difficult to handle in the image domain, 
\SYTnew{we propose to first reconstruct the 3D human model
from the given monocular video for both the source and target subjects
, and then adjust the pose of the target human model 
to match the source motion
(while maintaining the target person's body shape).}
Intrinsic geometric description of the deformed target is then projected back to 2D to form an image that reflects the 3D structure and serves as condition for subsequent generation. \sytminor{
We choose a few non-trivial eigenvectors of the Laplace matrix of the reconstructed mesh as the intrinsic geometry representation (dubbed ``Laplacian feature'' below), which is deformation-invariant and serves as a unique attribute of each vertex. 
Compared with other vertex-specific features, e.g., UV coordinates, the Laplacian feature is more suitable for the human generation task due to its spatially continuous embedding, implicitly encoded intrinsic geometry and higher feature dimensions. The boost of these characteristics to realistic human image generation is demonstrated in Section~\ref{sec:Ablation Study}.}
The projection of Laplacian feature along with the 2D pose figure extracted from each source image is used as a constraint during GAN-based image-to-image translation, 
to effectively maintain the structural characteristics of human body under different poses.

In addition, previous methods~\cite{chan18:every_dance_now, wang2018vid2vid} only use the appearance of the target person in the training process of pose-to-image translation. 
When an input pose is very different from any poses seen during the training process, such solutions might lead to blurry results. Observing that the source video frame corresponding to the input pose might contain reusable rich details (especially for the body parts like hands where the source and target subjects share some similarity), we intend to selectively transfer \SYTnew{geometry} details 
from real source frames to the synthesized video frames. This is achieved by our detail enhancement network.  {Figure~\ref{fig:teaser} shows representative motion transfer results with rich details}. 

We summarize our contributions as follows: 
\sytneww{1) We propose to use the
Laplacian features of the reconstructed parametric human model as 3D constraints 
for the human motion transfer task. 
Such a novel representation serves as a more intrinsic representation compared with other 3D features, e.g., UV coordinates adopted in~\cite{lwb2019}.}
2) We introduce the detail enhancement net (DE-Net), which utilizes the \sytneww{geometry} information 
from real source frames to enhance details in generated results. Extensive experiments show that our method outperforms the state-of-the-art methods~\cite{chan18:every_dance_now, wang2018vid2vid, lwb2019}.

\section{Related Work}
Over the last decades, motion transfer has been extensively studied due to its ability for fast video content production. Some early solutions have mainly revolved around realigning existing video footage according to the similarity to the 
desired pose \cite{Bregler1997VideoRD, Efros03}. However, it is not an easy task to find an accurate similarity measure for different actions of different subjects. Several other approaches have also attempted to address this problem in 3D, but they focus on the use of inverse kinematic solvers \cite{DBLP:conf/siggraph/LeeS99} and transfer motion between 3D \emph{skeletons} \cite{Hecker2008RealtimeMR}, whereas we consider using a reconstructed 3D body mesh to guide motion transfer in the image domain in order to provide 
much richer constraints.

Recently, the rapid advances of deep learning, especially generative adversarial networks (GANs) and their variations (e.g., cGAN \cite{mirza2014conditional}, CoGAN \cite{DBLP:conf/nips/LiuT16}, CycleGAN \cite{CycleGAN2017}, DiscoGAN \cite{DBLP:journals/corr/KimCKLK17}) have provided powerful tools 
for image-to-image translation,
which has yielded impressive results across a wide spectrum of synthesis tasks and shown 
its ability to synthesize visually pleasing images from conditional labels. For example, pix2pix \cite{isola2017image}, based on a conditional GAN framework, is one of the pioneering works. 
CycleGAN \cite{CycleGAN2017} further presents the idea of cycle consistency loss for learning to translate between two domains in the absence of paired images, and Recycle-GAN~\cite{bansal2018recycle} combines both spatial and temporal constraints for video retargeting tasks. Pix2pixHD~\cite{wang17:pix2pixHD} introduces a multi-scale conditional GAN to synthesize high-resolution images using both global and local generators, and vid2vid \cite{wang2018vid2vid} designs specific spatial and temporal adversarial constraints for video synthesis.

Based on these variants of GANs, a lot of approaches~\cite{Balakrishnan2018SynthesizingIO, chan18:every_dance_now,vunet2018, ma2017pose, ma2018disentangled} have been proposed for human motion transfer between two image domains. The key idea of these approaches is to decouple the pose information from an 
input image and use it as the input of a GAN network to generate a realistic 
image. For example, in~\cite{Balakrishnan2018SynthesizingIO}, an input image is separated into two parts: the foreground (or different body parts) and the background, and the final realistic image is generated by separate 
processing and cross fusion of the two parts. Chan \etal \cite{chan18:every_dance_now} extract the pose information with an off-the-shelf human pose detector OpenPose \cite{cao2018openpose} and use the pix2pixHD \cite{wang17:pix2pixHD} framework together with a specialized face 
GAN to learn a mapping from a 2D pose figure to an image. Neverova \etal\cite{neverova2018dense} adopt a similar idea but use the estimation of DensePose \cite{Guler2018DensePose} to guide image generation.
Wang \etal{} make a step further to adopt both OpenPose and DensePose in \cite{wang2018vid2vid}. However,  due to the lack of 3D semantic information 
, these approaches are highly sensitive to problems such as self-occlusions.

To solve the above problems, it is natural to add 3D information to the condition of generative networks. There are many robust 3D human mesh reconstruction methods (e.g., \cite{joo2018total, kanazawaHMR18, SMPL-X:2019, xiang2019monocular}), which can reconstruct a 3D human model with its corresponding pose from a single image or a video clip. Benefiting from these accurate and reliable 3D body reconstruction
techniques, we can study the issue of human motion transfer in a new perspective. Liu \etal\cite{lwb2019} present a novel warping strategy, which 
uses the projection of 3D models to tackle motion transfer, appearance transfer and novel view synthesis within a unified model. \SYTnew{However, due to the generative characteristic of their network for multi-person, it does not perform particularly well for the specific person.}


\section{Method}

\begin{figure*}[t]
\begin{center}
\includegraphics[width=0.95\textwidth]{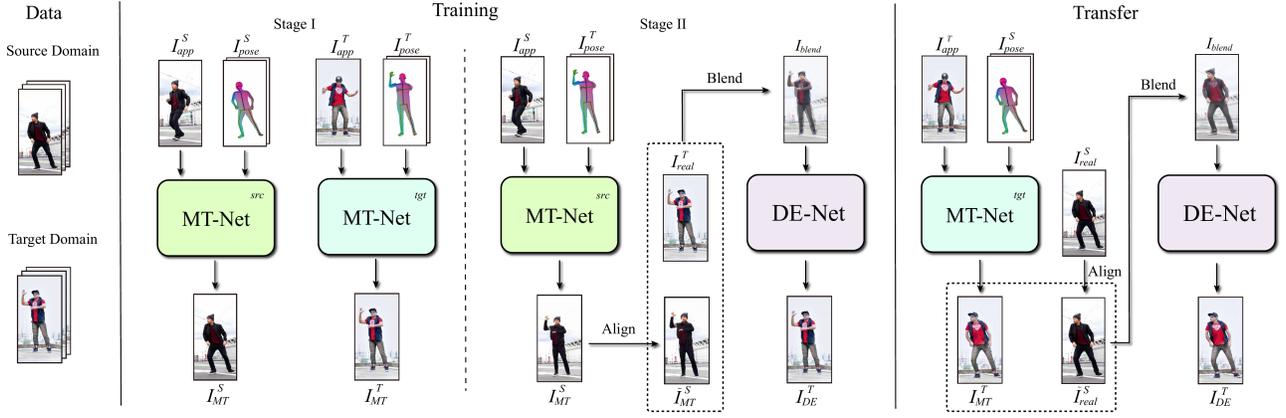}
\end{center}
\vspace{-3mm}
\caption{\SYTnew{The architecture of our pipeline. 
\textit{Training:} At Stage I, we train MT-Net(src) and MT-Net(tgt) separately within the corresponding domains. The MT-Net transfers the motion from temporally adjacent pose labels to the image domain represented by the appearance image.
At Stage II, we send a source appearance image $I_{app}^{\mathcal{S}}$ and adjacent target pose labels $I_{pose}^{\mathcal{T}}$ to MT-Net(src) to obtain a coarse source image $I_{MT}^{\mathcal{S}}$. After aligned to the target character's position and size, 
it is converted to the blending domain by blended with a corresponding real target frame. The DE-Net learns to translate the blended image 
to the target domain. \textit{Transfer:} Our system obtains 
a coarse transfer result $I_{MT}^{\mathcal{T}}$ with target appearance and the source pose labels by MT-Net(tgt), and converts $I_{MT}^{\mathcal{T}}$ to the  
blending domain by 
blending it with the aligned real source frame. Finally the DE-Net translates the blended image $I_{blend}$ to the target domain with details enhanced.}}
\vspace{-3mm}
\label{fig:pipeline}
\end{figure*}

We aim to generate a new video of a target person imitating the character movements in a source video, while keeping the structural integrity and detail features of the target subject as much as possible. To accomplish this, we use the mesh projection containing 3D information as the condition for the GAN, and introduce a detail enhancement mechanism to improve the details. 

\subsection{Overview}

\SYTnew{
We tackle this problem with a coarse-to-fine strategy. 
At the first stage, we design a Motion Transfer Net (MT-Net) to get initial motion transfer results
with the guidance of 3D information and a given appearance image.
Note that adjacent pose labels are used for temporal consistence. 
Intuitively, the MT-Net can be trained within the target domain and the initial transfer results can be obtained by simply replacing the target subject's pose label with the source character's.  However, such initial transfer results often exhibit blurring artifacts,  
especially in face and hands regions. To address this problem, we propose a Detail Enhancement Net} (DE-Net) accompanied with a complicated but effective training pipeline based on the observation that although being different in clothes or genders, source and target characters usually 
have similar structure in faces and hands, where we often suffer from blurring artifacts. It is thus possible to use the information in the source frames to enhance the synthesized details.
\sytneww{We first expound the main modules of our approach: 
pose label generation in Sec.~\ref{sec:pose_label}, motion transfer net (MT-Net) in Sec.~\ref{sec:MT-Net}, and detail enhancement net (DE-Net) in Sec.~\ref{sec:DE-Net}, followed by the training strategy in Sec.~\ref{sec:training_pipeline} and the optimization objective in Sec.~\ref{sec:full_objective}.}

\sytneww{\textbf{Notation}. We denote $\mathcal{S} = \{S_i\}$ as a set of source video frames, and $\mathcal{T} = \{T_j\}$ as a set of target frames. For $I_{sub}^{super}$, the subscript denotes the attribute of the image, while the superscript denotes the domain it belongs to.}


\subsection{Pose Label Generation} \label{sec:pose_label}
\begin{figure}[t]
\begin{center}
\includegraphics[width=0.9\linewidth]{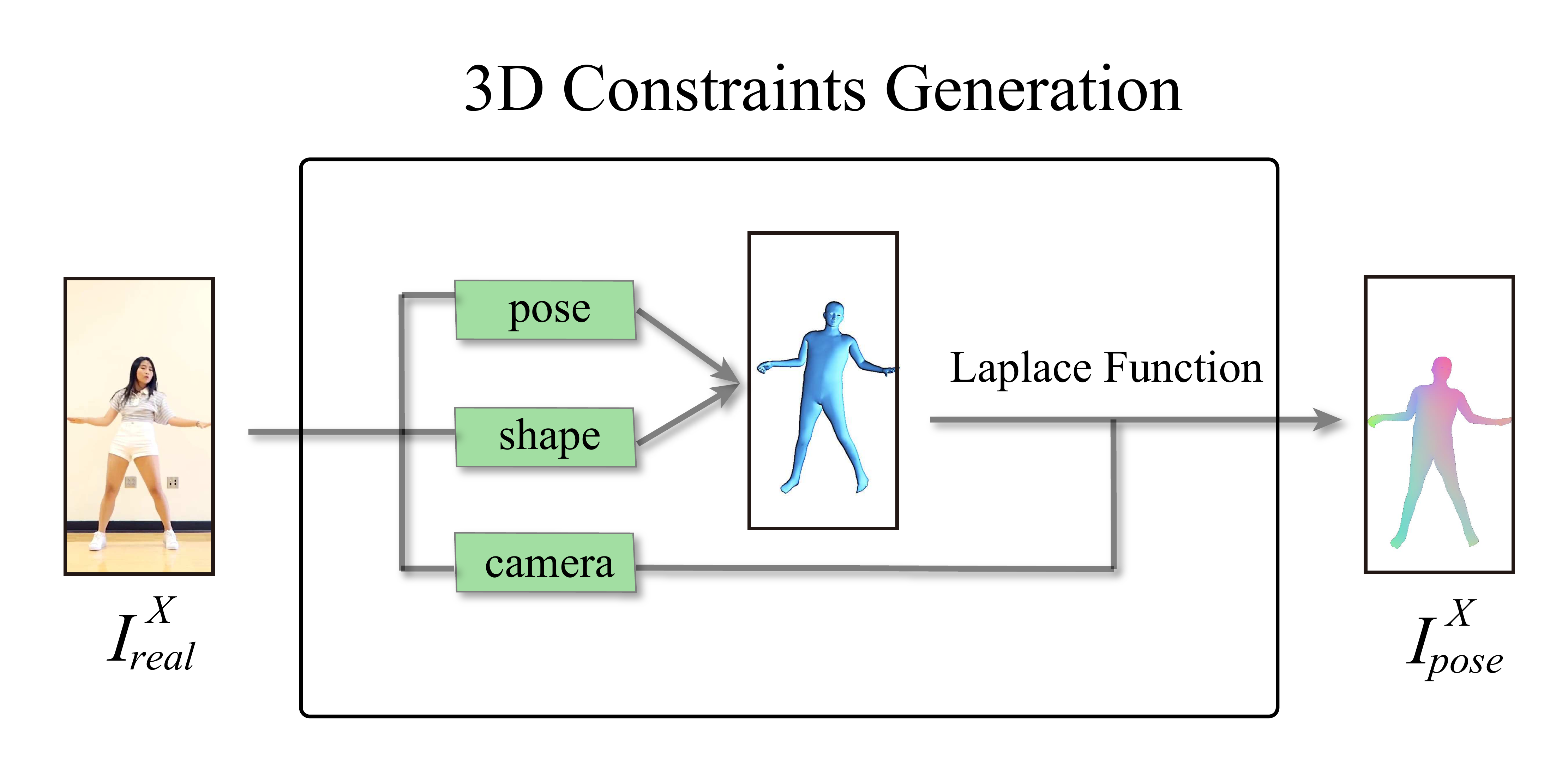}
\end{center}
\vspace{-6mm}
\caption{\hbmajor{Illustration} 
of the 3D constraints generation \hbmajor{process}. \SYT{We reconstruct a 3D mesh of \hbmajor{a source or target subject image $I_{real}^X$}, 
\hbmajor{associate} 
the three smallest eigenvalues of its Laplace matrix as intrinsic features (visualized in RGB color) \hbmajor{with the mesh vertices}, and project \hbmajor{the colored mesh} 
to form a 3D constraint image, denoted as $I_{pose}^X$.}}
\vspace{-3mm}
\label{fig:pose extractor}
\end{figure}

\SYTnew{Previous pose representations e.g. everybody~\cite{chan18:every_dance_now} often take the form of keypoints or skeleton, \hbmajor{and thus} 
are difficult to deal with self-occlusion and fail to maintain the structure integrity. Therefore we propose to utilize the 3D geometry information 
of the underlying subject to produce label images as the GAN condition to regularize the generative network. The process of our 3D constraints generation is shown in Fig.~\ref{fig:pose extractor}.} 

{\bf 3D Human Model Reconstruction.} We first extract the 3D body shape $\beta$ and pose $\theta$ information for both source and target videos using a state-of-the-art pre-trained 3D pose and shape estimator \cite{xiang2019monocular}. This leads to \hbmajor{two 3D deformable mesh models (for the source and target videos), }
including the details of body, face, and fingers. When transferring between \hbmajor{the} two domains, the 3D human models 
allow \hbmajor{easy} 
generation of a 3D mesh with the pose from one domain and shape from the other. The extracted deformable mesh sequences might exhibit temporal incoherence artifacts due to inevitable reconstruction errors. 
We alleviate this issue by simply applying temporal smoothing to \hbmajor{individual} mesh vertices, since our mesh sequences have the same connectivity.

{\bf Human Model Projection. } 
We project the 
reconstructed 3D human model onto 2D to obtain a label image, which will be used as the condition to guide the generator. The image should ideally contain intrinsic 3D information (invariant to pose changes) to guide the synthesis process such that \hbmajor{such information can be color-encoded: } a particular color correspond\hbmajor{ing} 
to a specific location on the human body. To achieve this, we propose to extract the three non-trivial eigenvectors corresponding to \hbmajor{the} three smallest eigenvalues \SYTnew{of the Laplace matrix of the reconstructed mesh} and consider them as a 3-channel color assigned to each vertex~\cite{meyer2003discrete}. \hbmajor{The colored mesh} 
is projected to 2D 
to form a 3D constraint image.

Note that although additional 3D information is available, 3D meshes extracted from 2D images may occasionally contain artifacts due to the inherent ambiguity. Therefore, we also adopt OpenPose \cite{cao2018openpose} to extract a 2D pose figure as part of the condition, which is less informative but more robust in the 2D space. 
Our label image therefore is 6-channel after combining both 2D and 3D constraints \SYT{actually}.
\SYT{We will \hbmajor{evaluate the impact} 
of these two conditions 
in the ablation study in Sec~\ref{sec:Ablation Study}}.

\subsection{Motion Transfer Net}
\label{sec:MT-Net}

\begin{figure*}[t]
\begin{center}
\includegraphics[width=0.8\linewidth]{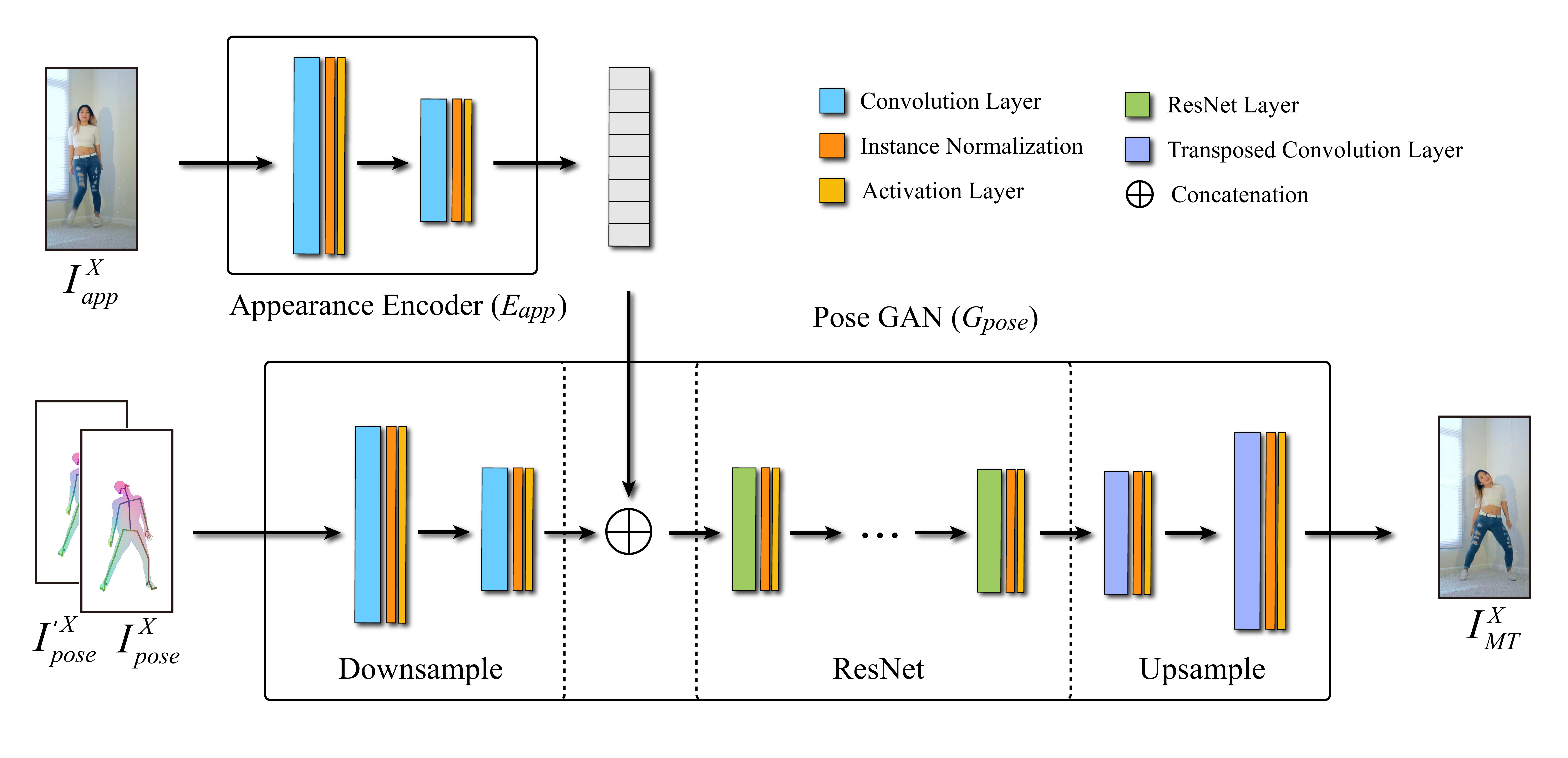}
\end{center}
\vspace{-9mm}
\caption{Architecture of the generative network. Appearance image $I_{app}^{\mathcal{X}}$ is randomly selected and sent to the Appearance Encoder (denoted as $E_{app}$) to obtain appearance features. Adjacent pose label images \hbmajor{including} $I_{pose}^{\mathcal{X}}$ (\hbmajor{for the} current frame) and $I_{pose}^{\mathcal{'X}}$ (\hbmajor{for the} previous frame) are sent to Pose GAN (denoted as $G_{pose}$) together with \hbmajor{the} appearance features to generate \hbmajor{a} 
reconstructed image $I_{MT}^{\mathcal{X}}$.}
\vspace{-3mm}
\label{fig:GAN}
\end{figure*}

\SYTnew{We learn the mapping from temporally adjacent label images 
and an appearance image to a realistic image 
by training the MT-Net, consisting of an Appearance Encoder and a Pose GAN (Figure~\ref{fig:GAN}). 
\sytneww{Specifically,  for $\mathcal{X} \in \{\mathcal{S}, \mathcal{T}\}$ let $I_{app}^{\mathcal{X}}$ and $I_{pose}^{\mathcal{X}}$ be the corresponding 
appearance image and the adjacent pose labels \sytneww{in the domain $\mathcal{X}$} 
(extracted from the current frame and the last frame), respectively, and we can obtain the reconstructed frame $I_{MT}^{\mathcal{X}}$.}
The Appearance Encoder encodes the appearance image $I_{app}^{\mathcal{X}}$ to a latent feature, and the Pose GAN produces an output image $I_{MT}^{\mathcal{X}}$ with the given appearance and pose/shape constraints. It is worth mentioning that in order to solve the problem of poor temporal continuity caused by single frame generation, similar to \cite{chan18:every_dance_now}, we involve adjacent frames
in $I_{pose}^{\mathcal{X}}$ to improve temporal coherence.} 

\textbf{Appearance Encoder $E_{app}$.}
The Appearance Encoder is a fully convolutional network that extracts appearance features from an input image $I_{app}^{\mathcal{X}}$, and the extracted feature 
is used as a condition for the Pose GAN. It takes randomly selected frames as input and outputs appearance features corresponding to that domain. 

\textbf{Pose GAN Generator $G_{pose}$.}
The Pose GAN is the main part of MT-Net,  which consists of three sub-modules: Downsampling, ResNet blocks, and Upsampling. It works on both label images and the appearance features extracted by the Appearance Encoder, and produces synthesized results with the corresponding pose and appearance. As shown in Fig.~\ref{fig:GAN}, the output of the Appearance Encoder is \SYTnew{concatenated to the pose features and the concatenated features are used as input to the ResNet blocks}.

\textbf{Pose GAN Discriminator $D^\mathcal{X}_{MT}$.}
We use the multi-scale discriminator presented in pix2pixHD \cite{wang17:pix2pixHD}\hbmajor{, since} 
discriminators of different scales can give the discrimination of images at different levels. In our method, we use two discriminators $D^\mathcal{S}_{MT}$ and $D^\mathcal{T}_{MT}$ to predict the probability of generated images belonging to the corresponding domains, each with 3 scales. 

The \hbmajor{generator and discriminator networks} 
drive each other: the generator learns to synthesize more realistic images conditioned on the input to fool the discriminator, while the discriminator in turn learns to discern the ``real'' images (ground truth) and ``fake'' (generated) images.

\sytneww{Note that MT-Net is not a generalized network that can be directly applied to any video character generation. Similar to pix2pixHD~\cite{wang17:pix2pixHD}, each trained MT-Net is bound to a specific video character. We condition the network on the additional appearance image since the common appearance information between frames helps the network converge faster and become more stable, as we demonstrate in the supplementary material.}

\SYTnew{As previously mentioned, in order to meet the need of the subsequent detail enhancement, we train the MT-Net in both \hbmajor{the} source domain and \hbmajor{the} target domain, denoted as MT-Net$^{src}$ and MT-Net$^{tgt}$, respectively. Note that these two networks share the same weights of \hbmajor{the} downsampling \hbmajor{and ResNet parts}, 
while \hbmajor{having separate weights for the upsamplling part.} 
}

\textbf{Temporal Smoothing.}
We use the time smoothing strategy in \cite{chan18:every_dance_now} to enhance the \hbmajor{temporal} continuity between adjacent generated frames. The generation of the current frame is not only related to the current label image $I_{label}$, but also related to the previous frame $I'_{label}$.

Therefore, 
our conditional GAN has the following objective:
\sytneww{
\begin{align}
\nonumber  \min_{\mbox{\tiny$\begin{array}{c}
  E_{app}, G_{pose}, \\  D^\mathcal{X}_{MT}\end{array}$}} \mathcal{L}^\mathcal{X}_{\text{MT}}(&I_{pose}^\mathcal{X}, I_{pose}^\mathcal{'X}, I_{real}^\mathcal{X}, I_{MT}^\mathcal{X}) =  \\
\nonumber &\mathbb{E}[\log D_{MT}^\mathcal{X}(I_{pose}^\mathcal{X}, I_{pose}^\mathcal{'X}, I_{real}^\mathcal{X})]  + \\
    &\mathbb{E}[\log(1-D_{MT}^\mathcal{X}(I_{pose}^\mathcal{X}, I_{pose}^\mathcal{'X}, I_{MT}^\mathcal{X})].
\label{equa:GAN}
\end{align}
}
Here the discriminator $D^\mathcal{X}_{MT}$ takes the pose labels and images in the domain $\mathcal{X}$ as input, and classifies them to real images (from the training set), or fake images ($I_{MT}$ generated by the Pose GAN).

\subsection{Detail Enhancement Net}
\label{sec:DE-Net}
\begin{figure}[t]
\begin{center}
\includegraphics[width=\linewidth]{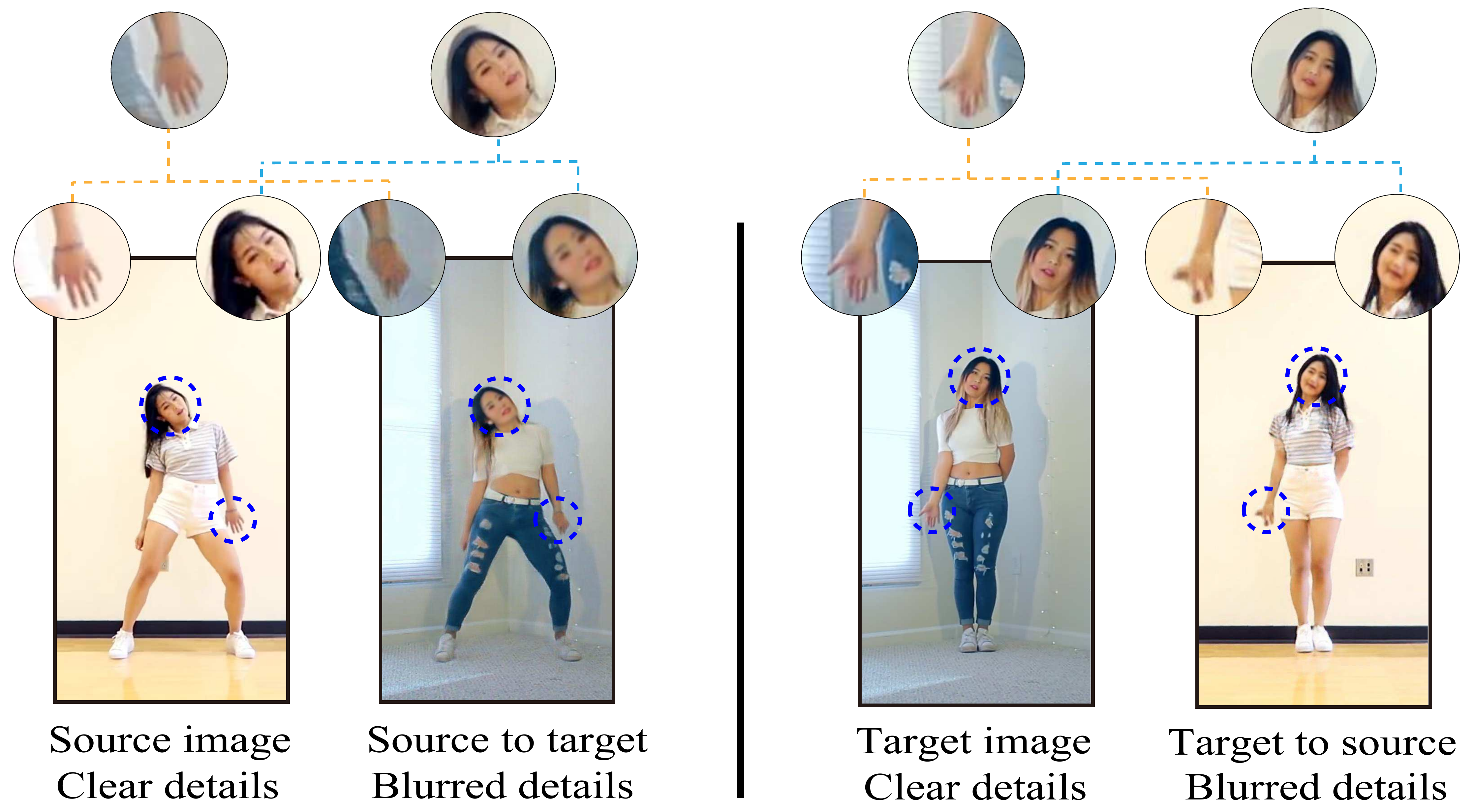}
\end{center}
\vspace{-5mm}
\caption{
\SYTnew{Comparison of a source image $I_{real}^{\mathcal{S}}$ and a source-to-target transfer result $I_{MT}^{\mathcal{T}}$, as well as a target image $I_{real}^{\mathcal{T}}$ and a target-to-source transfer result $I_{MT}^{\mathcal{S}}$. The upper circle\hbmajor{s show} 
the blending results in hands and face regions. The \hbmajor{blending} results \hbmajor{are of a similar style} 
for the two opposite blending modes.}}
\vspace{-3mm}
\label{fig:comparison}
\end{figure}

Through the first stage of training, we can obtain initial transfer results which, however, suffer from blurring artifacts.
Here we design the DE-Net for the recovery 
of blurred details.
Specifically, we aim to enhance details in the initial source-to-target 
transfer results $I_{MT}^{\mathcal{T}}$ with the help of the information from source frames. However, it is intractable for a neural network to extract desired details from the source domain and transfer them to the target domain. We address this problem based on the following observations:  
1) blurred regions concentrate in the hands and face regions of initial transfer results for both source-to-target and target-to-source; 2) hands and faces between different characters always have similar patterns, 
hence the blending of blurred regions with real frames can introduce reasonable structure information
3) the blending of source frames with blurred source-to-target transfer results has a similar style to
the blending of target frames with blurred target-to-source transfer results, as illustrated in Fig.~\ref{fig:comparison}.
Note that there are real target frames with the same poses as the target-to-source transfer results.  
This motivates us to train the DE-Net to learn the transfer 
from the blending domain to the target domain with supervision. The learned mapping can be used to enhance details for the source-to-target transfer results in the transfer stage.


{\bf Alignment and Blending.} Note that in different videos, subjects might have different builds or positions relative to the camera. In such cases, before blending we need to align the source frame ($I_{MT}^{\mathcal{S}}$ or $I_{real}^{\mathcal{S}}$) in accordance with the target by applying the transformation calculated from the reconstructed meshes with the source and target parameters (i.e., pose, body shape, and position). As we will show in Figure~\ref{fig:scale}, this step is effective in improving the details and avoiding artifacts. More details are included in the supplementary materials.
\SYTnew{We blend the images from the source and target domains linearly with a learnable parameter $\theta$.  Note that $\theta$ is updated during the training process and fixed in the inference(transfer) stage.
}


\begin{figure*}[t]
\begin{center}
\includegraphics[width=0.7\linewidth]{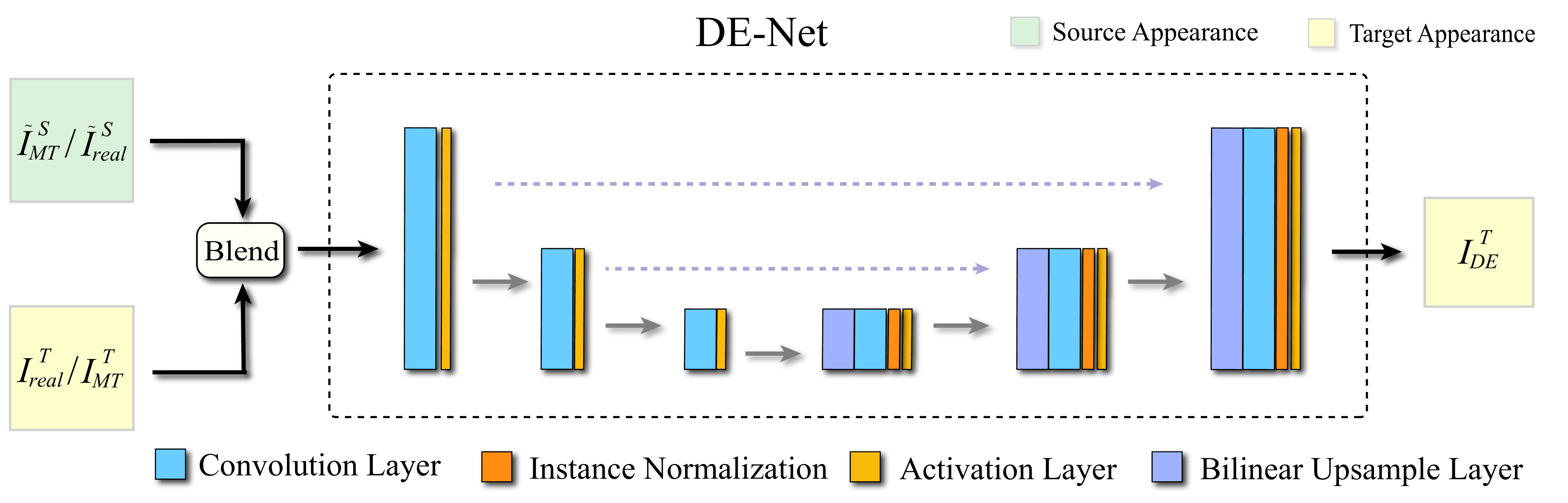}
\end{center}
\vspace{-5mm}
\caption{
\SYT{Architecture of Detail Enhancement Net (DE-Net). The main part of our DE-Net is a U-net, which takes the linear blending of paired data $(\widetilde{I}_{MT}^{\mathcal{S}}, I_{real}^{T})$ or $(\widetilde{I}_{real}^S, I_{MT}^T)$ as input, and synthesizes a target image $I^T_{DE}$ with details enhanced.}}
\vspace{-3mm}
\label{fig:Refine}
\end{figure*}

\sytneww{\bf Blending Domain to Target Domain.} The purpose of our DE-Net is to transfer the image from the blending domain to the target domain, while preserving useful details and eliminating redundant information from source frames.
It is based on the GAN framework, 
where the generator $G_{DE}$ is a U-net for synthesizing
images in the target domain with clear details, as illustrated in Figure \ref{fig:Refine}. 
The discriminator $D_{DE}$ discerns
 the ``real'' images (ground truth) and ``fake'' images (synthesized by $G_{DE}$).
 In the training stage, we use the blending of image pair $(I_{MT}^{\mathcal{S}}, I_{real}^{\mathcal{T}})$ as input and train DE-Net in a supervised manner with $I_{real}^{\mathcal{T}}$ as ground truth. The use of blended images instead of concatenation avoids the output overfitting to $I_{real}^{\mathcal{T}}$. \sytneww{We have verified the advantage of our blending manner in the supplementary.} 
We optimize the DE-Net by minimizing the following objective:
\vspace{-1.5mm}
\sytneww{
\begin{align}
\nonumber \min_{\mbox{\tiny$\begin{array}{c}
  G_{DE}, D_{DE}\end{array}$}} \mathcal{L}_{\texttt{DE}}(I_{pose}^{\mathcal{T}},& I_{real}^{\mathcal{T}}, I_{DE}^{\mathcal{T}}) = \mathbb{E}[\log D_{DE}(I_{pose}^{\mathcal{T}},  I_{real}^{\mathcal{T}})] \\
&  + \mathbb{E} [1 - \log D_{DE}(I_{pose}^{\mathcal{T}}, I_{DE}^{\mathcal{T}})],
\label{equa:refine}
\end{align}
}
where $I_{pose}^{\mathcal{T}}$ is the corresponding pose label.

In the transfer stage, we use the source-to-target transfer result $I_{MT}^{T}$ and the corresponding source image $I_{real}^{\mathcal{S}}$ to obtain enhanced transfer results.

\subsection{Data Flow in Different Phases}
\label{sec:training_pipeline}

\sytneww{We re-interpret the overall data flow in this subsection, including the training and transfer phases, 
as illustrated in Fig.~\ref{fig:pipeline}. The training part can be divided into two stages, i.e., the within-domain pre-training of MT-Net and the training of DE-Net.}


{\bf Within-Domain Pre-Training of MT-Net.} 
\SYTnew{To stabilize the training process, we first pre-train MT-Net using within-domain samples for both the source and target domains \sytneww{separately, namely the training of MT-Net$^{src}$ and MT-Net$^{tgt}$}. \sytminor{We use the superscript to distinguish them since MT-Net contains different parameters for different characters.}  To be specific, 
\sytneww{in the $\mathcal{X}$ domain the MT-Net$^{\mathcal{X}}$ is trained to map the appearance image $I_{app}^{\mathcal{X}}$ and adjacent pose labels $I_{pose}^{\mathcal{X}}$ to the reconstructed source frame.} Note that for each training $I_{app}^{\mathcal{X}}$ is randomly selected from video frames and fixed during the training process.} 
\sytneww{We have demonstrated the effect of appearance image selection in the supplementary material.}

{\bf Training of DE-Net.}
\sytneww{
Essentially, DE-Net is trained to learn the mapping from the blending domain to the target domain for detail extraction and preservation with the real target character frames as supervision.}
\SYTnew{As illustrated in Stage II of Fig.~\ref{fig:pipeline}, in order to train DE-Net we generate an initially transferred image $I_{MT}^{\mathcal{S}}$ of \textit{the source character driven by a given target pose} $I_{pose}^{\mathcal{T}}$. This process seems to running 
counter to the final objective \sytneww{(i.e., generating a target image from source character pose $I_{pose}^{\mathcal{S}}$)}, however, it sets the stage for the supervised 
training of DE-Net} \sytneww{with $I_{real}^{\mathcal{T}}$ as the ground truth.}

{\bf Transfer.} 
\SYTnew{Our \textit{transfer (or inference) pipeline} is straightforward. By sending $I_{app}^{\mathcal{T}}$ and $I_{pose}^{\mathcal{S}}$ to MT-Net, we can obtain the initial transfer result $I_{MT}^{\mathcal{T}}$. DE-Net takes the blending of $I_{MT}^{\mathcal{T}}$ and its corresponding source frame as input, and outputs the final result with details enhanced.}



\subsection{Full Objective}
\label{sec:full_objective}
The training of our network is divided into two stages. First we train the Motion Transfer Net. 
The full objective has the following form, containing 
adversarial loss $\mathcal{L}^{\mathcal{X}}_{\text{MT}}$, perceptual loss $\mathcal{L}_{P}$, and discriminator feature-matching loss $\mathcal{L}_{FM}$: 
\sytneww{
\begin{align}
\nonumber   \min \limits_{E_{app},G_{pose}}& \sum \limits_{\mathcal{X} \in {\mathcal{S}, \mathcal{T}}} ((\max \limits_{D^\mathcal{X}_{MT}} \mathcal{L}^{\mathcal{X}}_{\text{MT}}(I_{pose}^\mathcal{X}, I_{pose}^\mathcal{'X}, I_{real}^\mathcal{X}, I_{MT}^\mathcal{X})) \\
\nonumber    &+ \lambda_P \mathcal{L}_P(I_{MT}^{\mathcal{X}}, I_{real}^\mathcal{X}) \\ 
 &+ \lambda_{FM}\mathcal{L}_{FM}(I_{pose}^\mathcal{X}, I_{pose}^\mathcal{'X}, I_{real}^\mathcal{X}, I_{MT}^\mathcal{X}).
\end{align}
}

Here, $\mathcal{L}^{\mathcal{X}}_{\text{MT}}$ is defined in Eq.~\ref{equa:GAN}. The perceptual loss $\mathcal{L}_{P}$ regularizes a generated result $I_{MT}^{\mathcal{X}}$ to be close to the ground truth $I_{real}^{\mathcal{X}}$ in the VGG-19 \cite{simonyan2014very} feature space, defined as
\begin{equation}\label{eq:Lp}
\mathcal{L}_P(I_1, I_2)  = ||\text{VGG}(I_{1}) - \text{VGG}(I_{2})||_1.
\end{equation}
The discriminator feature-matching loss $\mathcal{L}_{FM}$ 
presented in pix2pixHD \cite{wang17:pix2pixHD} 
similarly regularizes the output using intermediate results of the discriminator, and is calculated as
\sytneww{
\begin{align}\label{eq:Lfm}
\nonumber \mathcal{L}_{FM}(&I_{pose}^\mathcal{X}, I_{pose}^\mathcal{'X}, I_{real}^\mathcal{X}, I_{MT}^\mathcal{X}) = \\ &
\mathbb{E} \sum \limits_{i=1} \limits^T \frac{1}{N_i}[||D_k^{(i)}(s,x)-D_k^{(i)}(s,G(s))||_1],
\end{align}}%
where \sytneww{$D$ denotes the discriminator $D_{MT}^{\mathcal{X}}$}, $T$ is the number of layers, $N_i$ is the number of elements in the $i$-th layer, and $k$ is the index of discriminators in the multi-scale architecture. $s$ is the condition of cGAN and $x$ is the corresponding ground truth.

The DE-Net is optimized with the following objective
\vspace{-2mm}
\sytneww{
\begin{align}
\nonumber    &\min \limits_{G_{DE}}((\max \limits_ {D_{DE}} \mathcal{L}_{\text{DE}}(I_{pose}^{\mathcal{T}},I_{real}^{\mathcal{T}}, I_{DE}^{\mathcal{T}})) + \lambda_P \mathcal{L}_P(I_{DE}^{\mathcal{T}}, I_{real}^{\mathcal{T}}) + \\
& \lambda_{FM}\mathcal{L}_{FM}(I_{pose}^\mathcal{X}, I_{pose}^\mathcal{'X}, I_{real}^\mathcal{X}, I_{MT}^\mathcal{X})) + \lambda_{reg} ||\theta - 0.5||_2.
\end{align}
}
Here $\mathcal{L}_{\text{DE}}$ is defined in Eq.~\ref{equa:refine}. The perceptual and discriminator feature-matching losses are defined in Eqs.~\ref{eq:Lp} and \ref{eq:Lfm}, respectively. \SYTnew{The last term regularizes the blending coefficient to avoid the overfitting to target frames.}


\begin{figure}[t]
\centering
    \includegraphics[width=\linewidth]{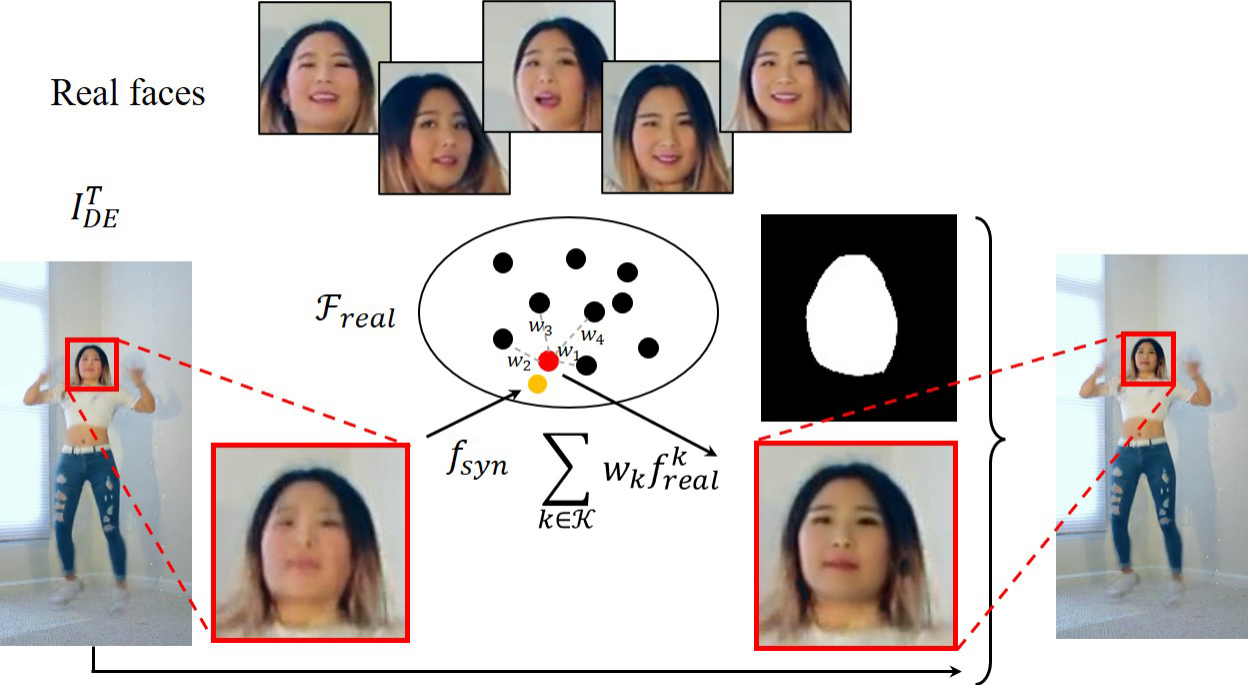}
    \vspace{-6mm}
    \caption{\sytneww{The illustration of post-processing facial enhancement. The cropped facial part $f_{syn}$ of the synthesized frame $I^T_{DE}$ is embedded into the StyleGAN3 latent space together with real faces $\mathcal{F}_{real} = \{f_{real}^{i}\}$. Through a retrieval-and-interpolation strategy, the latent embedding is refined to get an enhanced face, 
    which is finally fused to the synthesized frame with a predicted facial mask.}}
    \vspace{-3mm}
    \label{fig:face_enhancement_illustration}
\end{figure}

\vspace{-5mm}
\sytneww{
\subsection{Facial Enhancement}
To further improve the quality of synthesized human frames, we add a post-processing step 
to enhance 
facial details. We assume that the StyleGAN3~\cite{Karras2021AliasFreeGA} latent 
codes of the same person's facial images constitute a low-dimensional, locally-linear manifold. Based on this assumption, the synthesized facial part can be enhanced with a retrieval-and-interpolation approach by projecting it into the StyleGAN3 latent space, as similarly done in DeepFaceDrawing~\cite{chenDeepFaceDrawing2020}. 
Fig.~\ref{fig:face_enhancement_illustration} illustrates this idea.
Specifically, the facial images with different expressions and angles are first cropped from real frames, and then encoded to the StyleGAN3 latent space with e4e~\cite{Tov2021DesigningAE}. These latent features are denoted as $\mathcal{F}_{real} = \{f_{real}^{i}\}$,  where the superscript $i$ indicates the $i$th feature. Given a synthesized frame of the same character, the facial part can be extracted and embedded as $f_{syn}$ with the same approach. We aim to project $f_{syn}$ to the manifold constructed by $\mathcal{F}_{real}$. We first find the $K$-nearest neighbors of $f_{syn}$ in $\mathcal{F}_{real}$ under the Euclidean space, whose indexes form 
a set $\mathcal{K}$. 
With the locally linear assumption, $f_{syn}$ can be projected as a 
linear combination of these neighbors with coefficients $w_k\  (k \in \mathcal{K})$ by minimizing the reconstruction error. This 
can be formulated 
as the following optimization problem
\begin{equation}
    \min || f_{syn} - \sum_{k \in \mathcal{K}} w_k f_{real}^{k} || \qquad s.t. \sum_{k \in \mathcal{K}} w_k = 1.
\end{equation}
The coefficients can be determined by solving a constrained least-squares problem. Accordingly, $f_{syn}$ can be refined as $\sum_{k \in \mathcal{K}} w_k f_{real}^{k}$, which is then fed into StyleGAN3 for more detailed facial part generation. Finally the enhanced facial image is fused to the synthesized frame ($I_{DE}^{\mathcal{T}}$) according to the detected facial mask.
We show the effect of facial enhancement in Fig.~\ref{fig:face_enhancement}}.

        

        


\section{Experiments}
We compare our method with state-of-the-art methods and ablation variants, both quantitatively and qualitatively.

\subsection{Setup}
{\bf Datasets.}
To evaluate the performance of our method, we collected three types of data: the dataset published by~\cite{chan18:every_dance_now}, a few single-dancer videos from YouTube and Bilibili, and 5 videos filmed by ourselves (2 with ordinary background and 3 with green screen). 
Each subject wears different clothes and performs different types of action such as freestyle dancing and stretching exercises. \sytminor{\hbminor{We obtained these subjects' consent to use their video data for our experiments.}} 
We cut out the start and end parts that contain no action,
crop and normalize each frame to $1024 \times 512$ resolution by simple scaling and translation. Note that each video in our dataset can serve as either a source or target video. 

\begin{figure*}[t]
\begin{center}
\includegraphics[scale=0.085]{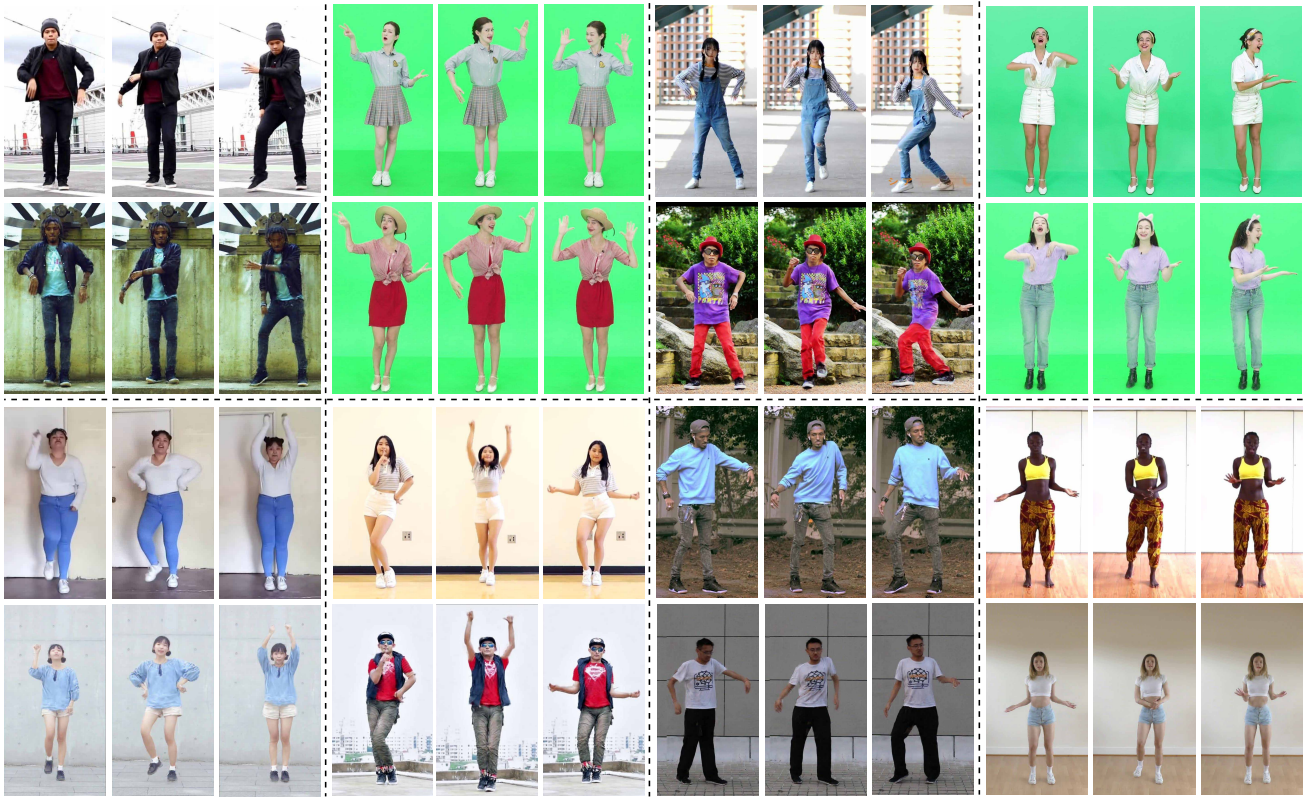}
\end{center}
\vspace{-3mm}
\caption{\hbmajor{Motion} transfer results. We show the generated frames of several subjects \SYT{with different genders, races, and builds}. In each group, the top row shows the source subject and the bottom row shows the generated target subject. \hbmajor{Please refer to the supplemental materials for synthesized videos.} }
\vspace{-3mm}
\label{fig:sample}
\end{figure*}

{\bf Implementation Details.} 
\label{sec:training}
We adopt a multi-stage training strategy in our method using Adam optimizer with \hbmajor{the} learning rate \hbmajor{of} 0.0001.
In the first stage, we pre-train the MT-Net for 20 epochs. In the next stage, the parameters of MT-Net are fixed and \hbmajor{the} DE-Net is trained individually for 10 epochs.
We set hyper-parameters $\lambda_{FM} = 10$ \hbmajor{and} 
$\lambda_P = 5$ for both stages and $\lambda_{reg} = 10$ for the second stage.  More details about MT-Net and DE-Net \hbmajor{training} are given in \hbmajor{the} supplementary material\hbmajor{s}.

{\bf Existing Methods.}
We compare our \hbmajor{method} 
with 
state-of-the-art methods \emph{vid2vid} \cite{wang2018vid2vid}, \emph{Everybody Dance Now} \cite{chan18:every_dance_now} and \emph{Liquid Warping GAN} \cite{lwb2019}, using \hbmajor{their} official implementation.

\begin{figure}[t]
\begin{center}
\includegraphics[width=0.9\linewidth]{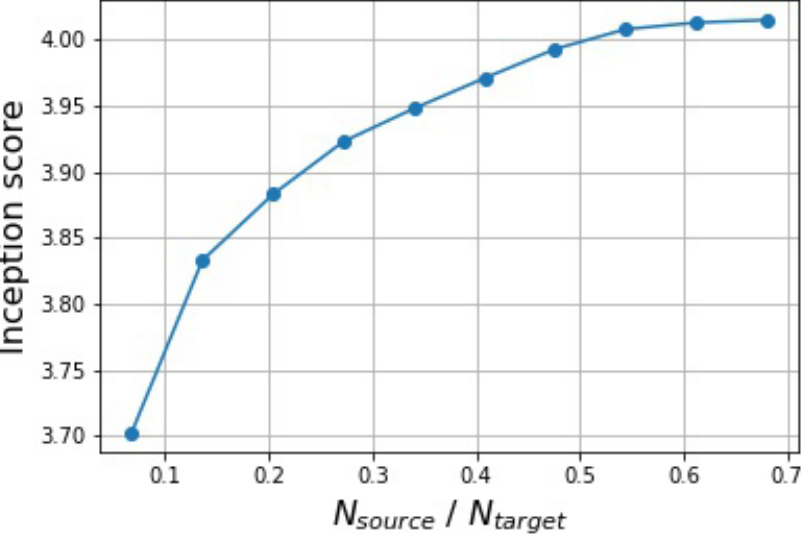}
\end{center}
\vspace{-5mm}
\caption{\SYT{The effect of \hbmajor{the} number of source frames on transfer results. We show the inception score of transfer results ($y$-axis) with respect to 
the ratio of source frame number to target ($x$-axis). For \hbmajor{the} inception score, \hbmajor{the} higher is \hbmajor{the} better.}
}
\vspace{-5mm}
\label{fig:loss}
\end{figure}

\begin{figure}[t]
\begin{center}
\vspace{-5mm}
\includegraphics[scale=0.24]{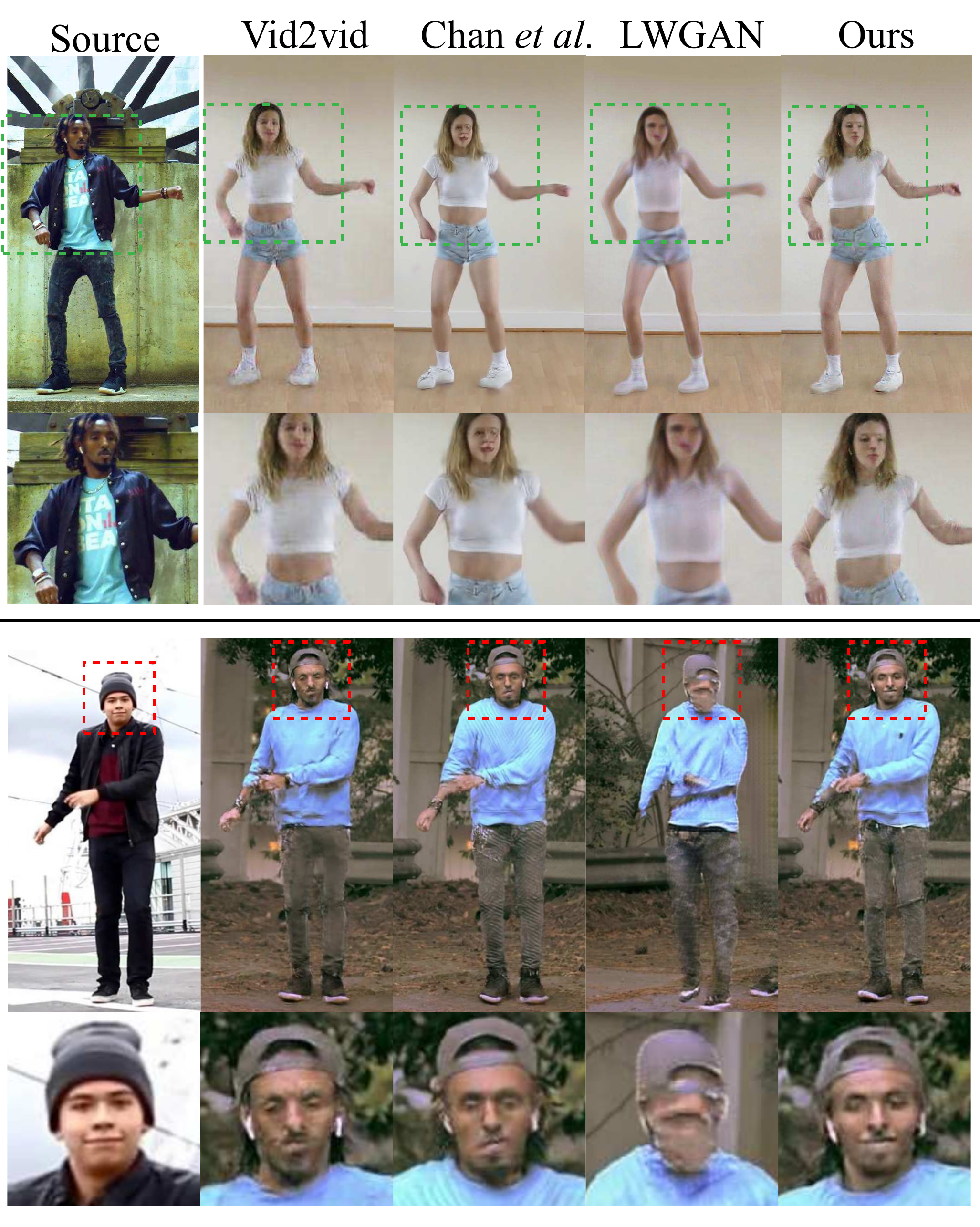}
\end{center}
\vspace{-5mm}
\caption{Comparison\hbmajor{s} with \hbmajor{the} state-of-the-art methods. We show the generated results by vid2vid, Everybody Dance Now \hbmajor{by Chan et al.}, Liquid Warping GAN \hbmajor{(LW-GAN),} and our method. Our method \hbmajor{produces richer and more realistic details}. 
}
\vspace{-5mm}
\label{fig:contrast}
\end{figure}

\begin{figure*}[t]
\begin{center}
\includegraphics[scale=0.12]{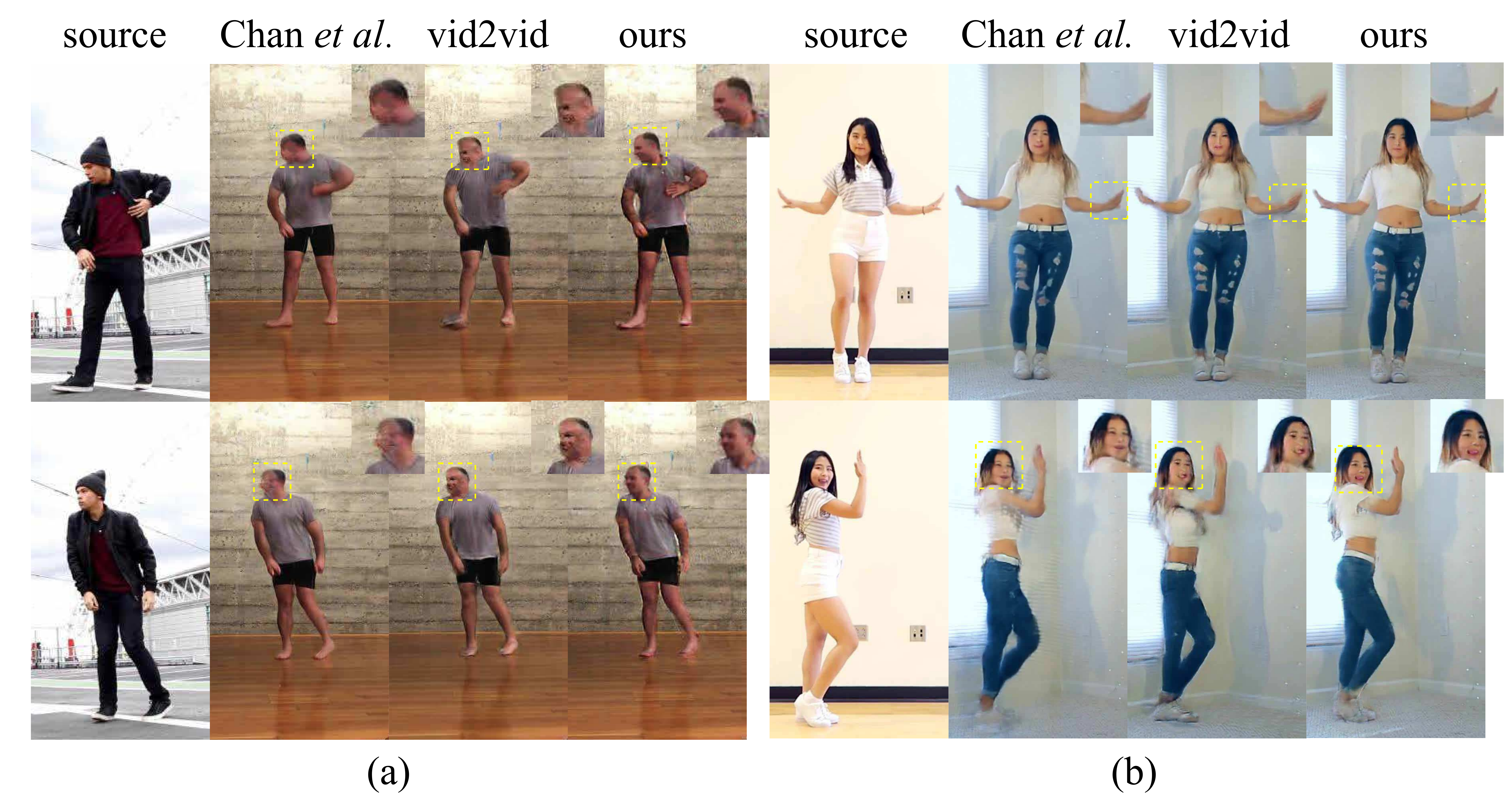}
\end{center}
\vspace{-5mm}
\caption{\SYT{We compare \hbmajor{our approach} with \hbmajor{the method by} Chan \etal~\cite{chan18:every_dance_now} and vid2vid~\cite{wang2018vid2vid} on the data published by~\cite{chan18:every_dance_now} (a) and the data used in~\cite{wang2018vid2vid} (b) for the sake of fairness. Our method \hbmajor{produces} 
more \hbmajor{realistic} 
results 
for occluded actions such as side faces and bending fingers.}}
\vspace{-5mm}
\label{fig:contrast_more}
\end{figure*}


\subsection{Quantitative Results}
{\bf Evaluation Metrics.}
We use objective metrics for quantitative evaluation under two different conditions: 1) To directly measure the quality of generated images, we perform self-transfer, in which the source and target are the same subject, and then use SSIM \cite{wang2004image} and learning-based perceptual similarity (LPIPS) \cite{zhang2018perceptual} to assess the similarity between source and target images. We split frames of each subject into \hbmajor{a training set} and \hbmajor{a}  \SYTnew{validation} set at the ratio of 8:2 for this evaluation. 2) We also evaluate the performance of cross-subject transfer, where the source and target are different subjects, using inception score \cite{salimans2016improved} and Fréchet Inception Distance (FID)~\cite{heusel2017gans} as metrics. It should be noted that we compute the FID score between the original and generated target images since there exists no ground truth for comparison in this case.
\sytneww{We also fine-tune the inception module~\cite{Szegedy2016RethinkingTI} with the Market-1501 dataset~\cite{zheng2015scalable} on the Person-ReID task, which serves as a more accurate feature extractor for synthesized human images due to the reduced domain gap, as demonstrated in~\cite{Li2019DenseIA}. We denote this metric as ``\textbf{IS-ReID}''.}
We exclude the green screen dataset in quantitative evaluation to focus on more challenging cases.

The metrics mentioned above are all based on single frames, which cannot reflect the smoothness of generated image sequences. The effect of mesh filtering in time series can be observed in the video results and quantitatively measured by the user study in Sec~\ref{sec:user evaluation}. 

\subsubsection{Comparison with Existing Methods.}
Comparison results with the state-of-the-art methods are reported in Table \ref{table1}.
It can be found that our method performs better than others, in both self-transfer and cross-subject transfer.

\begin{table}[t]
\caption{Quantitative comparisons with the state-of-the-art methods on the dance dataset. \SYTnew{It can be seen that our method outperforms SOTA in both self-transfer and corss-transfer condition.}}
\vspace{-6mm}
\begin{center}
\begin{tabular}{l|l|llll}
\hline
\multicolumn{2}{c|}{\multirow{2}{*}{Metric}} & \multicolumn{4}{c}{Method} \\ \cline{3-6} 
\multicolumn{2}{c|}{} & \footnotesize{vid2vid} & \footnotesize{Chan \etal} & \footnotesize{LW-GAN} &ours \\ \hline
\multirow{2}{*}{\begin{tabular}[c]{@{}l@{}}Self-\\ trans\end{tabular}} & \small{SSIM $\uparrow$} & 0.781 & 0.836 & 0.790 & \sytneww{\textbf{0.891}} \\ \cline{2-6} 
 & \small{LPIPS $\downarrow$} & 0.096 & 0.067 & 0.106 & \sytneww{\textbf{0.039}} \\ \hline
\multirow{2}{*}{\begin{tabular}[c]{@{}l@{}}\small{Cross-}\\ trans\end{tabular}} & \sytneww{\small{IS-ReID $\uparrow$}} & \sytneww{3.568}& \sytneww{3.794} & \sytneww{3.274} & \sytneww{\textbf{4.015}} \\ \cline{2-6} 
 & \small{FID $\downarrow$} & 59.98 & 57.02 & 81.20 & \sytneww{\textbf{51.26}} \\ \hline
\end{tabular}
\end{center}
\vspace{-4mm}
\label{table1}
\end{table}


\subsubsection{Ablation Study}
\label{sec:Ablation Study}
We perform an ablation study to verify the impact of each \hbmajor{key} component of our model, including 
using 3D constraints (``\textbf{3D}'') and DE-Net (``\textbf{DE}''). Our full pipeline is indicated as ``Full''. Note that \SYTnew{``\textbf{Full (ave)}'' means blending images from \hbmajor{the} source and target domain in an average way.}

Table \ref{table2} shows the results of the ablation study. It is obvious that our full proposed framework performs better than \hbmajor{its} 
variants. 
Both 3D constraints and DE-Net are able to enhance the results.
\SYT{Although there is no explicit 3D loss, the Laplace projection of 3D meshes effectively defines the shape and geometry
information and serves as a condition of Pose GAN.
The 3D constraints plays an important role in generation, as 
shown in \textbf{MT (3D)} and \textbf{MT (2D)}.
The score\hbmajor{s} of \textbf{MT(2D+3D)} 
show the complementarity of 2D and 3D conditions on this task.}
The comparison \hbmajor{between} 
\textbf{Full} and \textbf{MT(2D+3D)} (or \hbmajor{between} \textbf{MT(2D)+DE} and \textbf{MT(2D)}) proves the \hbmajor{positive effect} 
of DE-Net.
\SYTnew{The difference of \hbmajor{the} blending approaches is reflected in the ``\textbf{Full (ave)}'' and ``\textbf{Full}'', showing the advantage of the learned blending parameter.}

Furthermore, we can observe that \hbmajor{the} scores of self-transfer between \textbf{MT(2D)} and \textbf{MT(2D+3D)} (or \hbmajor{between} \textbf{MT(2D)+DE} and \textbf{Full}) are similar. This is mainly because source and target subjects share the same body shape in self-transfer, 
thus somewhat limiting the effectiveness (and necessity) of 3D constraints. \hbmajor{In contrast,} 
the scores of cross-subject transfer indicate 
the important role \hbmajor{of the} 3D information 
on transfer between different subjects with different shapes.


\begin{table*}[h]
\vspace{-2mm}
\caption{Ablation study. \SYTnew{Our approach achieves the best scores \hbmajor{among the compared variants}. 
}}
\vspace{-6mm}
\begin{center}
\resizebox{0.7\linewidth}{!}{
\begin{tabular}{l|l|cccccc}
\hline
\multicolumn{2}{c|}{\multirow{2}{*}{Metric}} & \multicolumn{6}{c}{Method} \\ \cline{3-8} 
\multicolumn{2}{c|}{} & {MT(2D)} &
\begin{tabular}[c]{@{}l@{}}{\SYT{MT(3D)}}\end{tabular} & \begin{tabular}[c]{@{}l@{}}{MT(2D+3D)}\end{tabular} & \begin{tabular}[c]{@{}l@{}}{MT(2D)+DE}\end{tabular} &
\begin{tabular}[c]{@{}l@{}}{Full (ave)}\end{tabular}& {Full} \\ \hline
\multirow{2}{*}{\begin{tabular}[c]{@{}l@{}}{Self}-\\ {trans}\end{tabular}} & {SSIM $\uparrow$} & 0.828 & 0.831 & 0.856  & 0.877 & 0.887 & \sytneww{\textbf{0.891}}\\ \cline{2-8} 
 & {LPIPS $\downarrow$} & 0.064 & 0.063 & 0.058 & 0.043 & 0.041 & \sytneww{\textbf{0.039}} \\ \hline
\multirow{2}{*}{\begin{tabular}[c]{@{}l@{}}{Cross}-\\ {trans}\end{tabular}} & {\sytneww{IS-ReID} $\uparrow$} & \sytneww{3.523} & \sytneww{3.632} & \sytneww{3.731} & \sytneww{3.587} & \sytneww{3.954} &   \sytneww{\textbf{4.015}} \\ \cline{2-8} 
 & {FID $\downarrow$} & 58.62 & 56.68 & 55.77 & 57.56  & 53.76 & \sytneww{\textbf{51.26}} \\ \hline
\end{tabular}}
\end{center}
\label{table2}
\end{table*}

\begin{figure*}[t]
\begin{center}
\includegraphics[width=\linewidth]{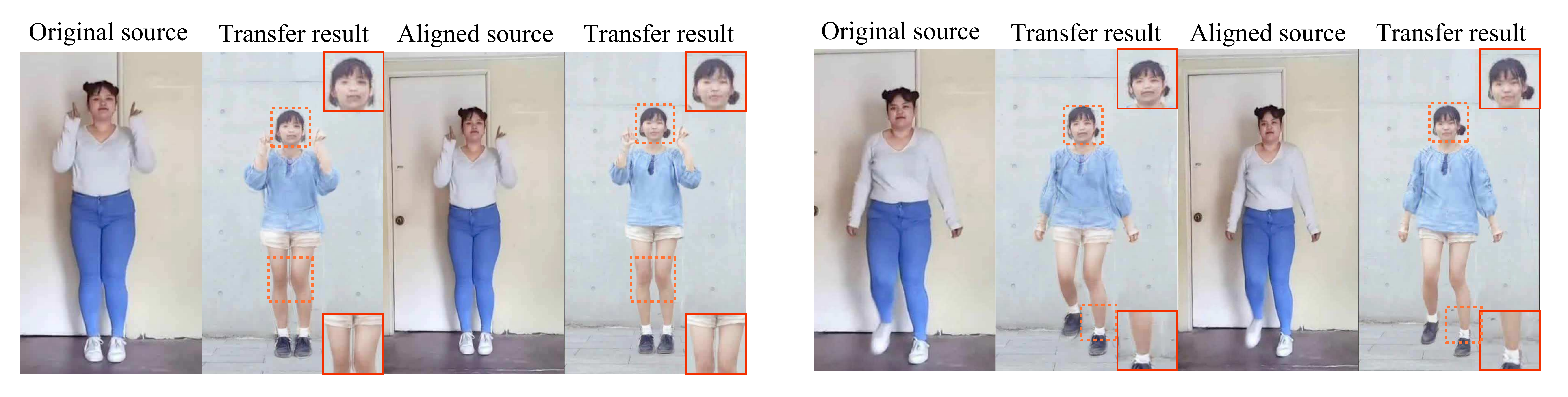}
\end{center}
\vspace{-5mm}
\caption{\SYT{Effects of aligning transformation for DE-Net. It can be seen that the absence of transformation will make DE-Net unable to match source and target characters accurately, resulting in unclear results with artifacts and body shape changes.}}
\vspace{-5mm}
\label{fig:scale}
\end{figure*}

\begin{figure}[t]
\begin{center}
\subfigure[Effect of 3D constraints. With 3D constraints, the results are improved in movements with occlusion such as bending legs (left) and structural integrity such as profile synthesis (right).]{\includegraphics[scale=0.11]{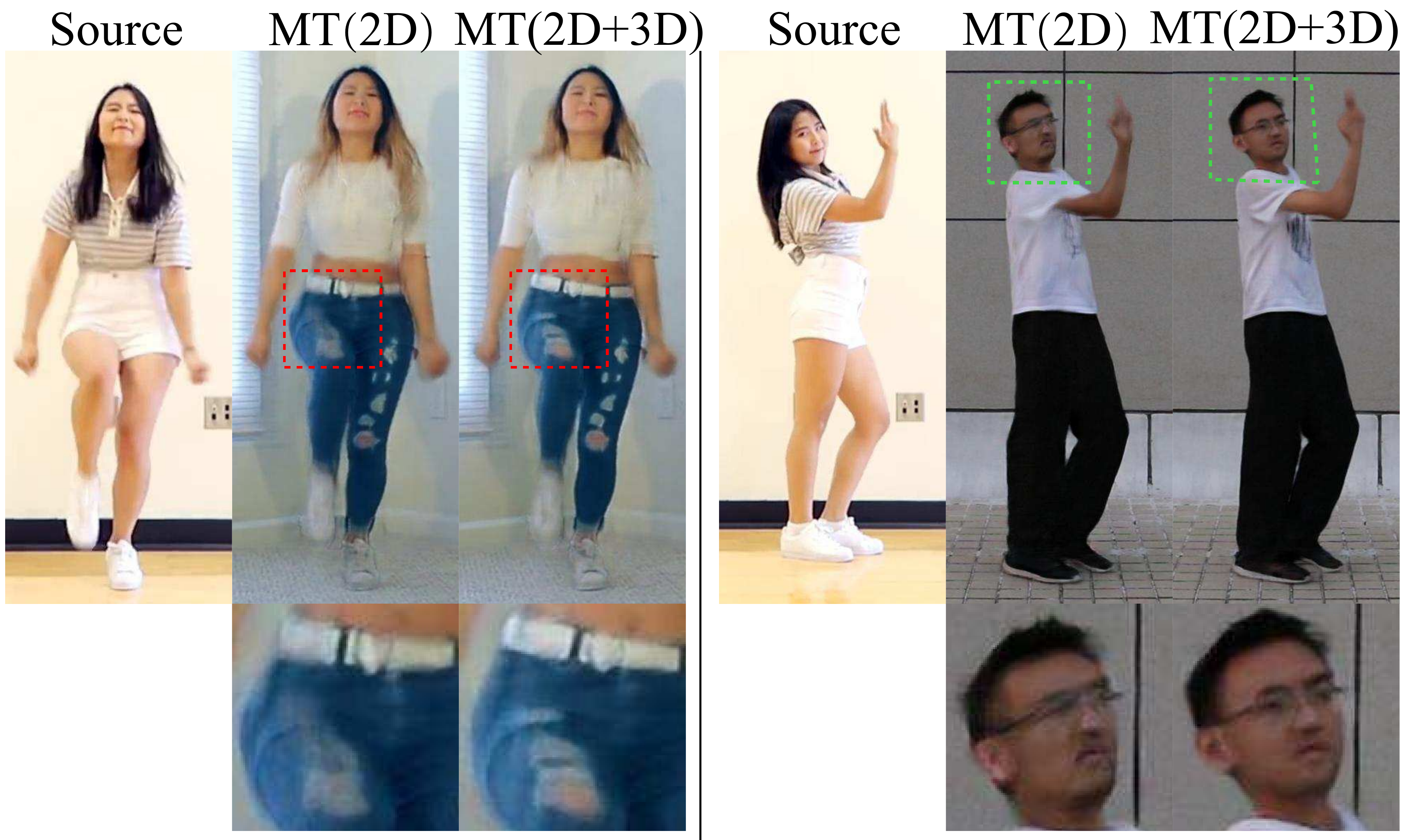}\label{fig:ablation1}}
\subfigure[Effect of the DE-Net. The DE-Net shows superior results in the generation of details in the face (left) and hand (right) areas. ]{\includegraphics[scale=0.11]{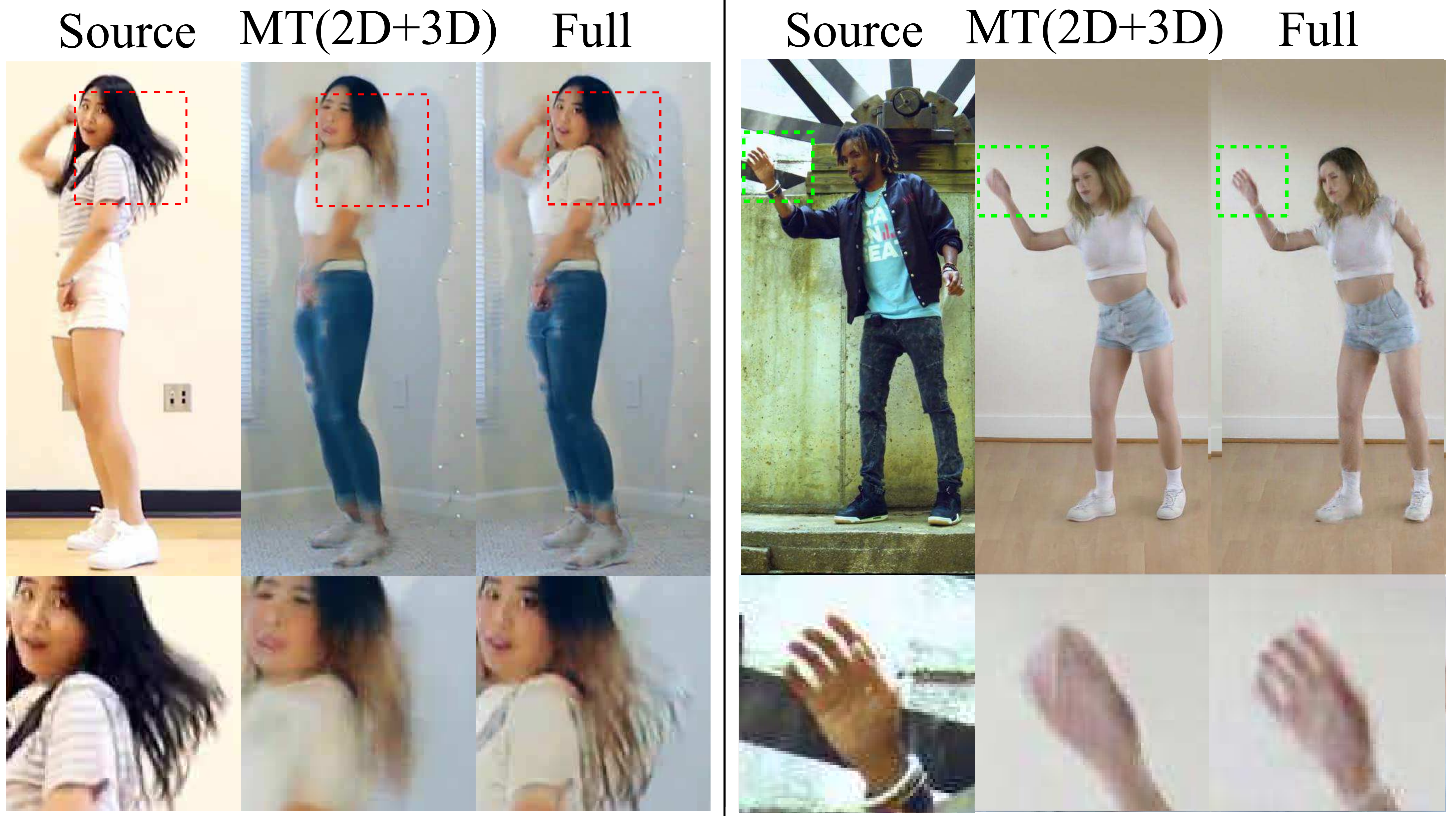}\label{fig:ablation2}}
\end{center}
\vspace{-5mm}
\caption{Visual comparison for the ablation study. We show the generated results of different conditions set in the ablation study.}
\vspace{-3mm}
\label{fig:ablation}
\end{figure}

\begin{figure}[t]
    \centering
    \setlength{\fboxrule}{0.5pt}
    \setlength{\fboxsep}{-0.01cm}
    \begin{spacing}{1}
    \begin{tabular}{p{0.005\linewidth}<{\centering}p{0.12\linewidth}<{\centering}p{0.12\linewidth}<{\centering}p{0.12\linewidth}<{\centering}p{0.12\linewidth}<{\centering}p{0.12\linewidth}<{\centering}p{0.12\linewidth}<{\centering}}

    & UV~\cite{lwb2019} & Depth~\cite{Prokudin2021SMPLpixNA}  & Ours & UV~\cite{lwb2019} & Depth~\cite{Prokudin2021SMPLpixNA}  & Ours \\
    \vspace{-22mm} \rotatebox{90}{3D pose label} &
    \includegraphics[width=1.15\linewidth]{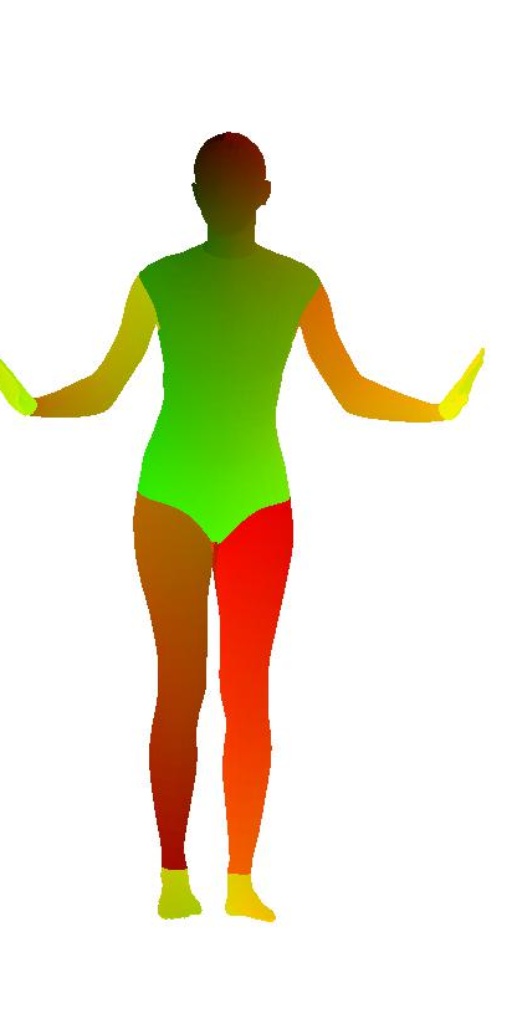} &
    \includegraphics[width=1.15\linewidth]{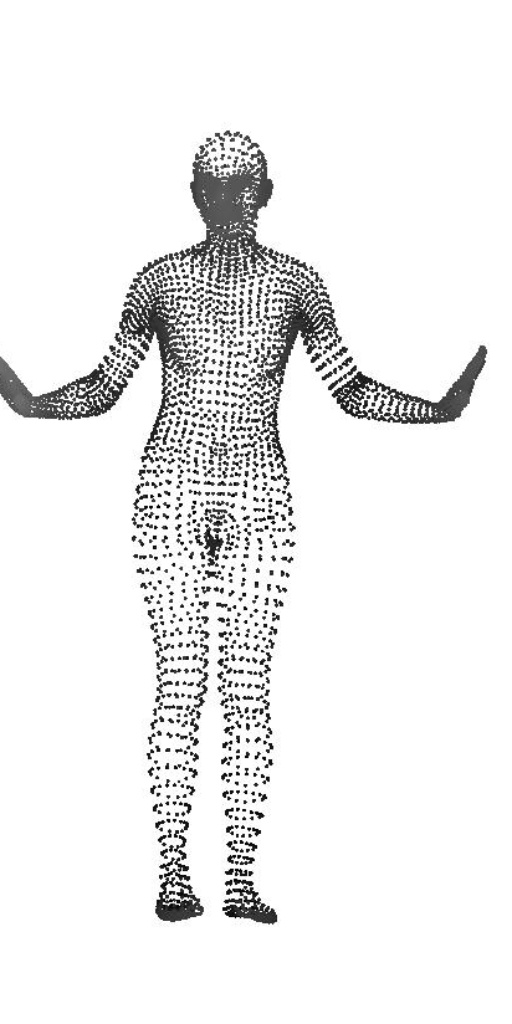} &
    \includegraphics[width=1.15\linewidth]{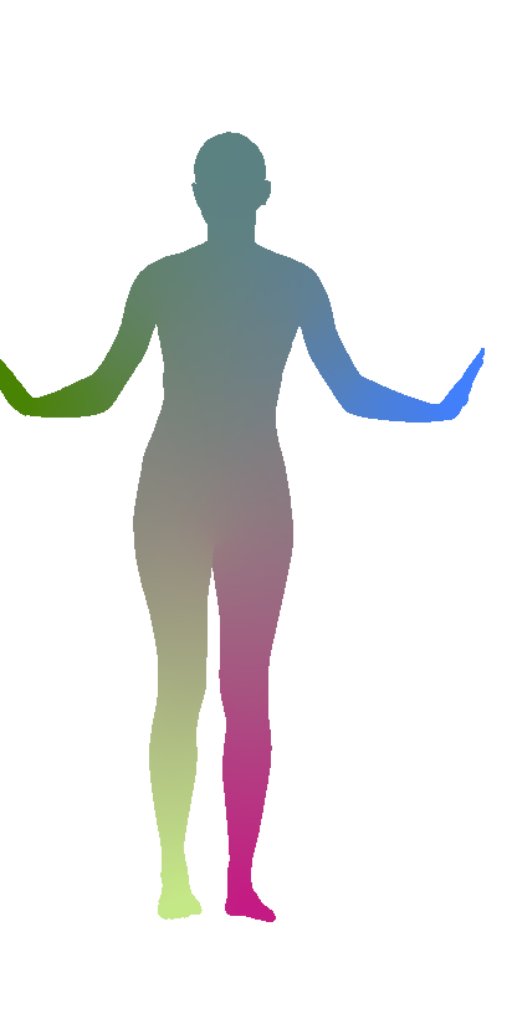} &
    \includegraphics[width=1.15\linewidth]{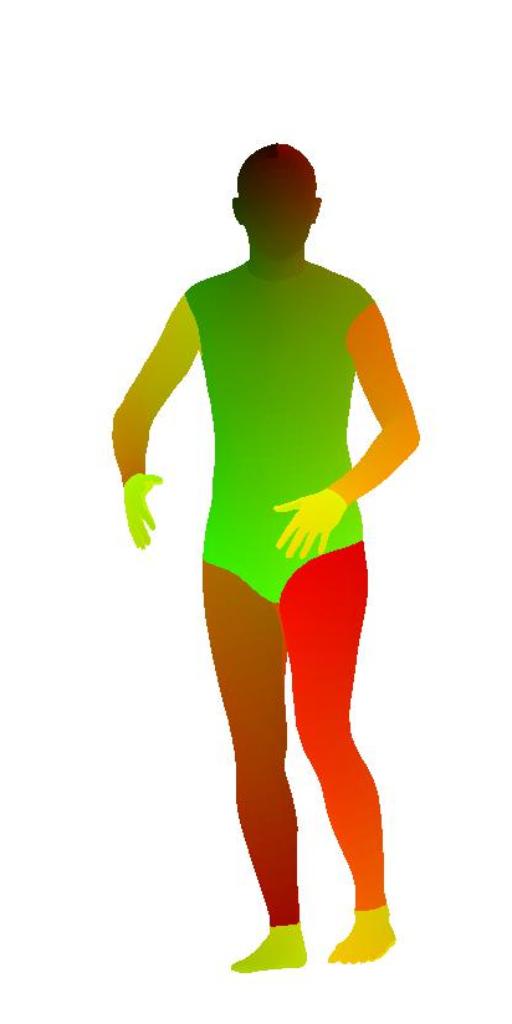} &
    \includegraphics[width=1.15\linewidth]{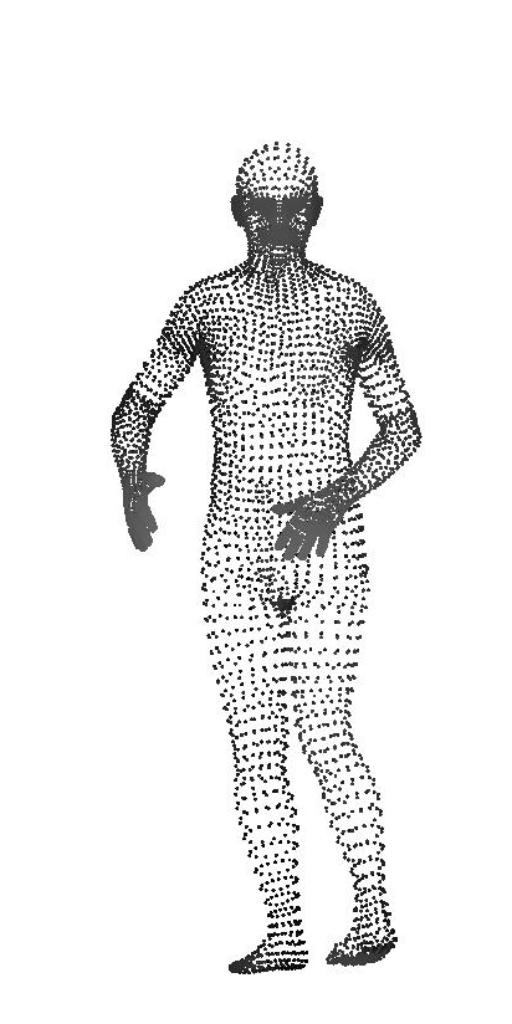} &
    \includegraphics[width=1.15\linewidth]{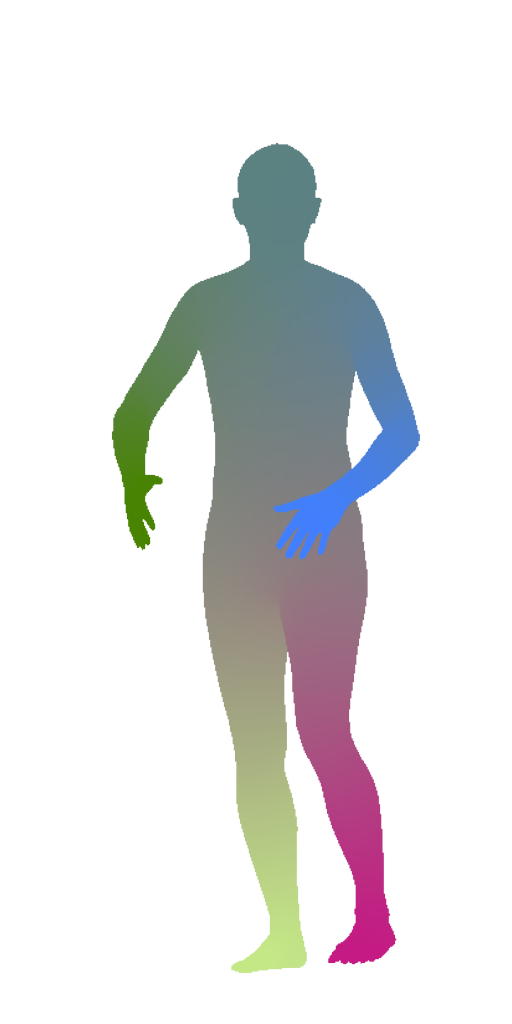}
    \\
    \specialrule{0em}{0pt}{-16pt} 
    \\
    
    \vspace{-22mm} \rotatebox{90}{MT-Net result} &
    \includegraphics[width=1.15\linewidth]{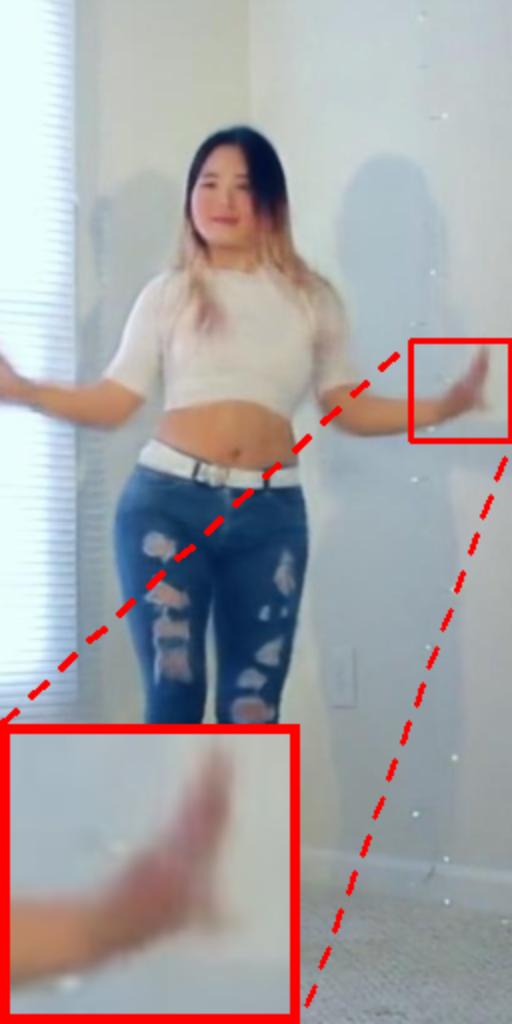} &
    \includegraphics[width=1.15\linewidth]{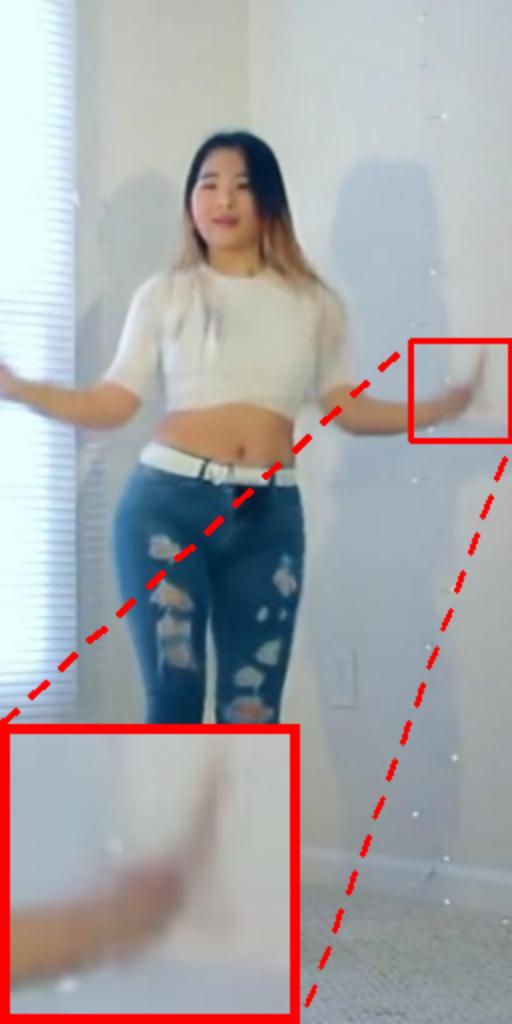} &
    \includegraphics[width=1.15\linewidth]{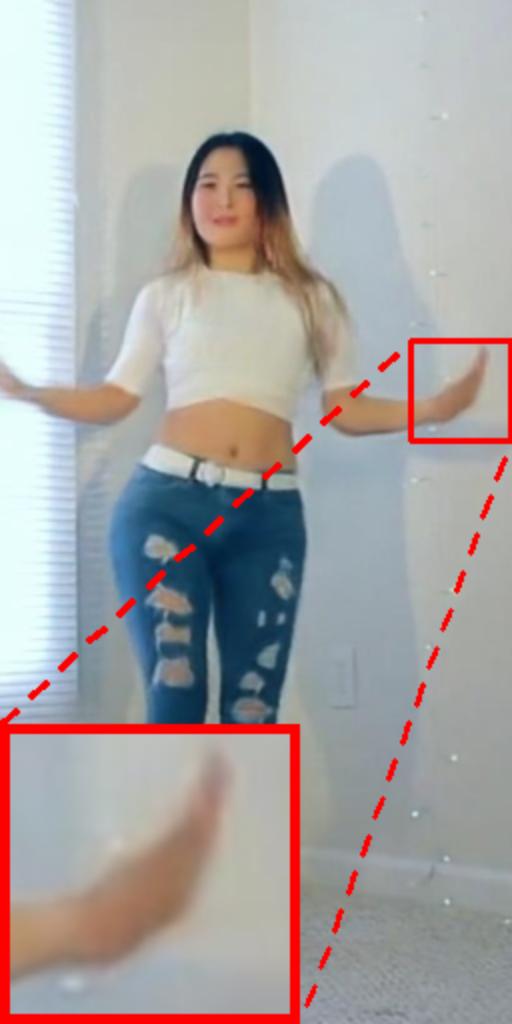} &
    \includegraphics[width=1.15\linewidth]{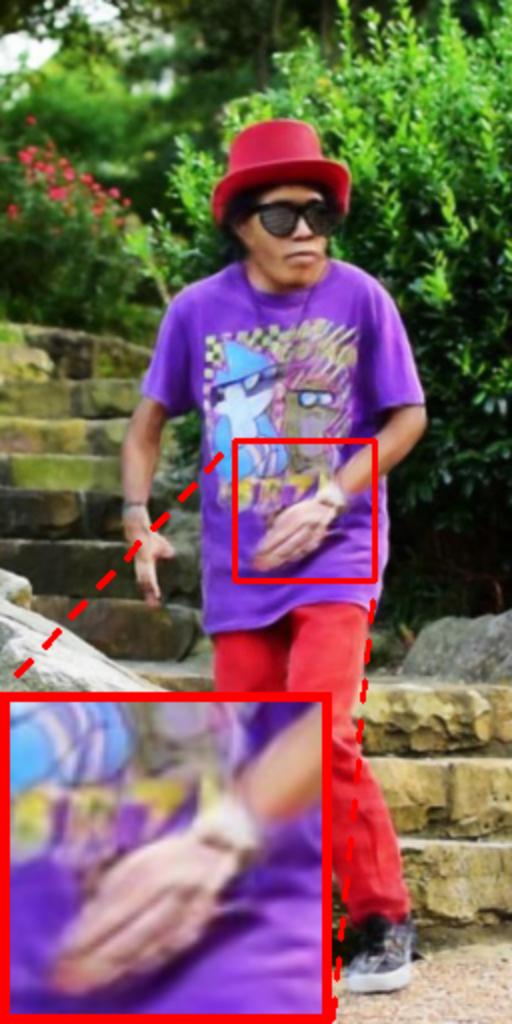} &
    \includegraphics[width=1.15\linewidth]{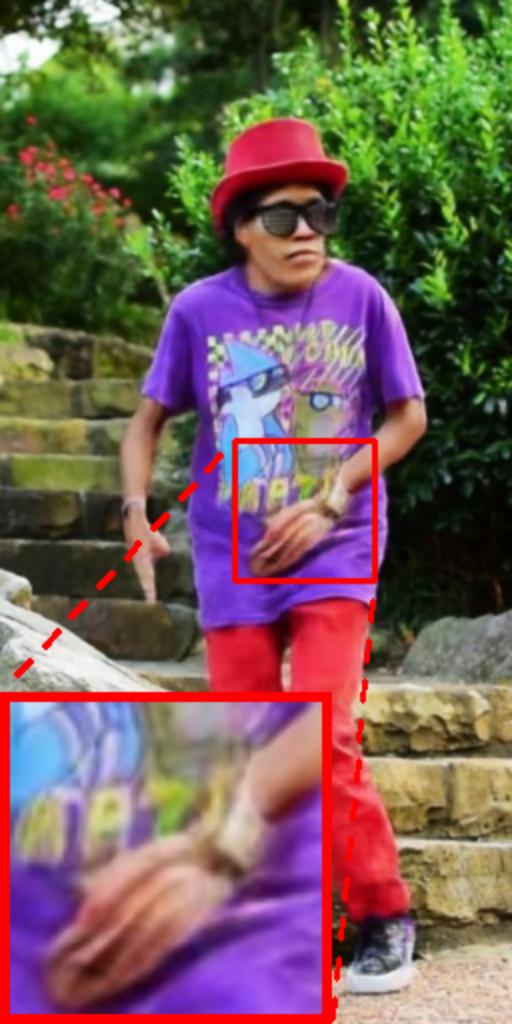} &
    \includegraphics[width=1.15\linewidth]{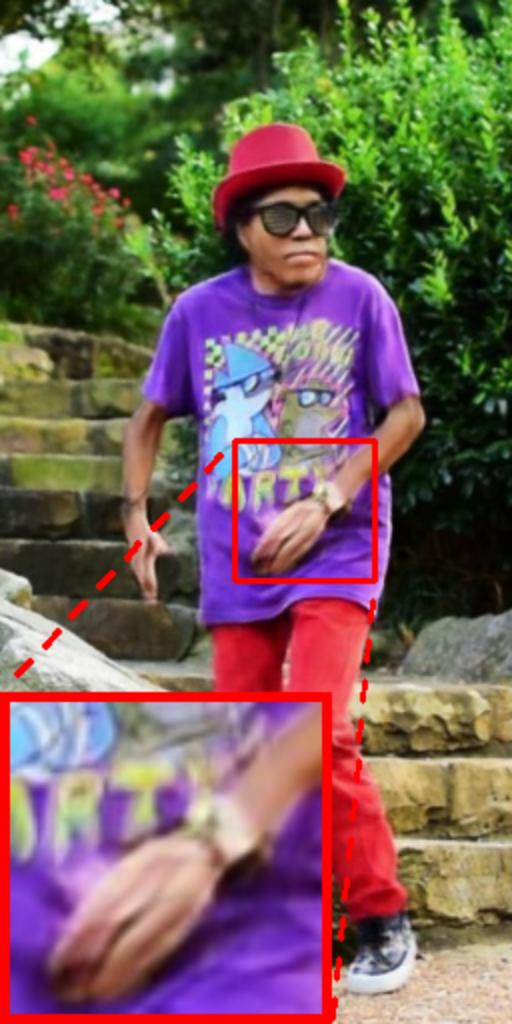}
    \\
    
    \vspace{-22mm} \rotatebox{90}{DE-Net result} &
    \includegraphics[width=1.15\linewidth]{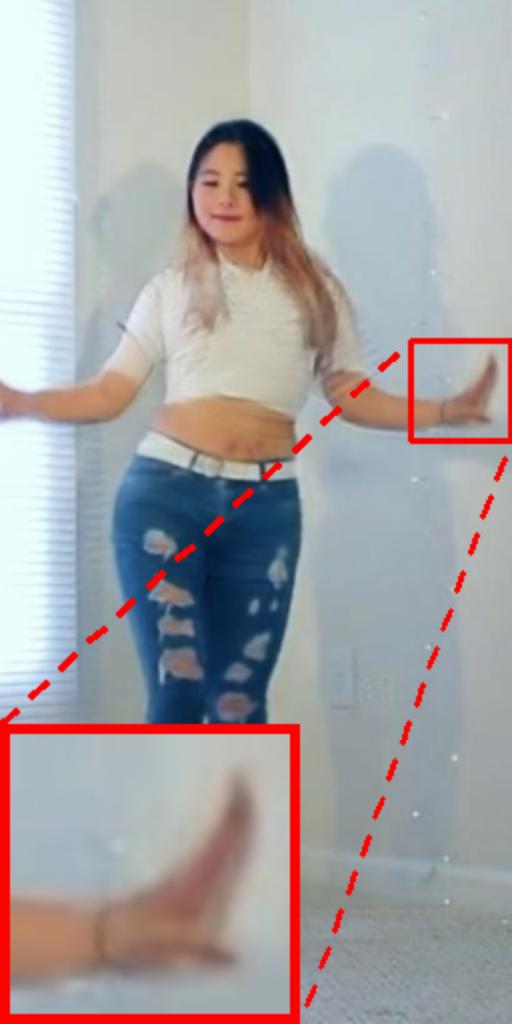} &
    \includegraphics[width=1.15\linewidth]{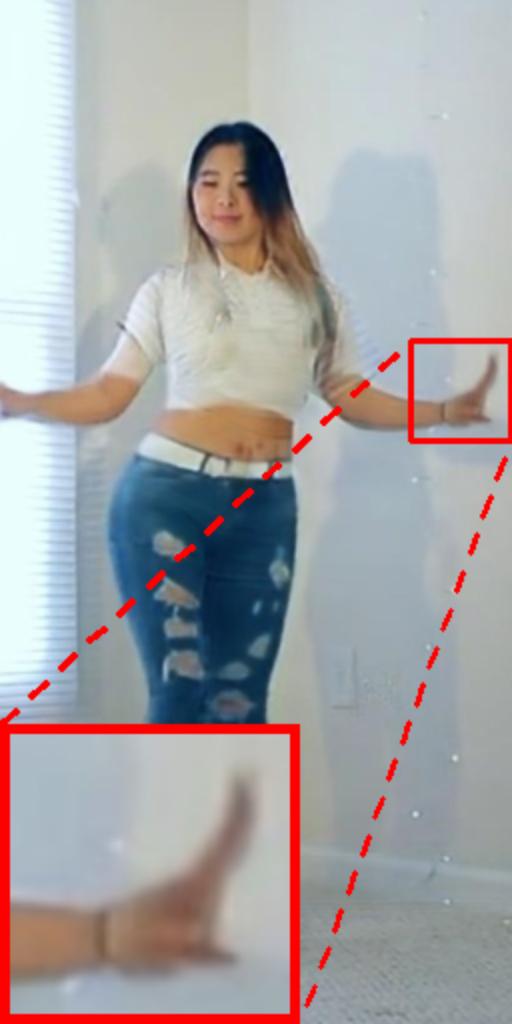} &
    \includegraphics[width=1.15\linewidth]{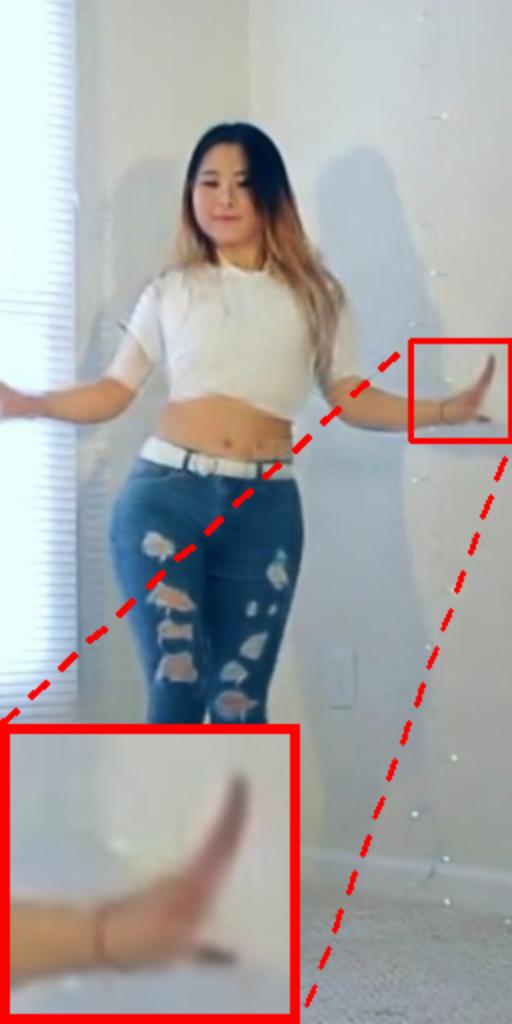} &
    \includegraphics[width=1.15\linewidth]{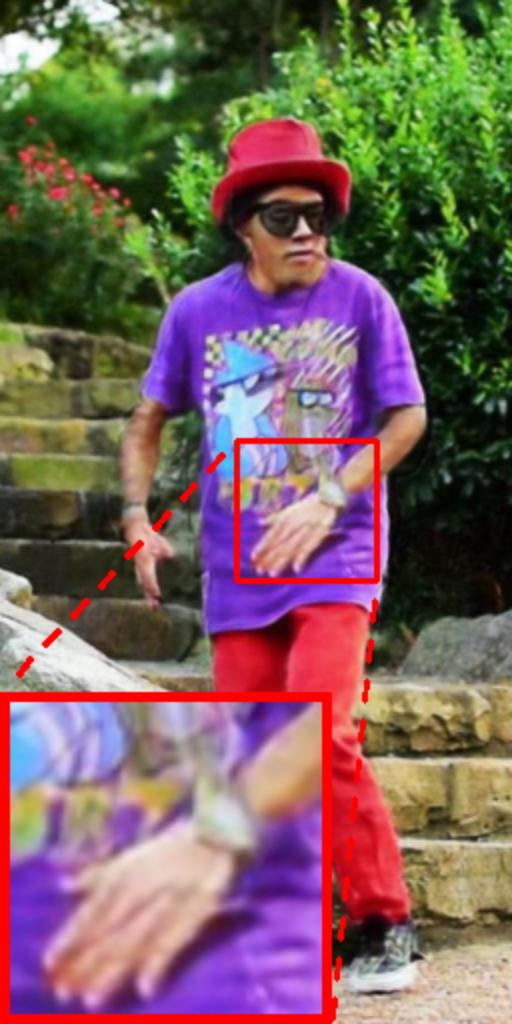} &
    \includegraphics[width=1.15\linewidth]{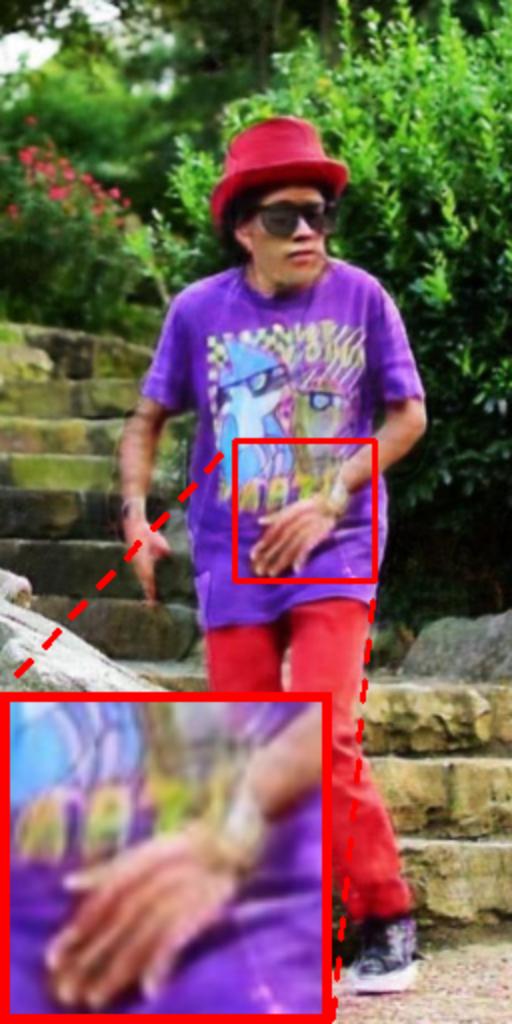} &
    \includegraphics[width=1.15\linewidth]{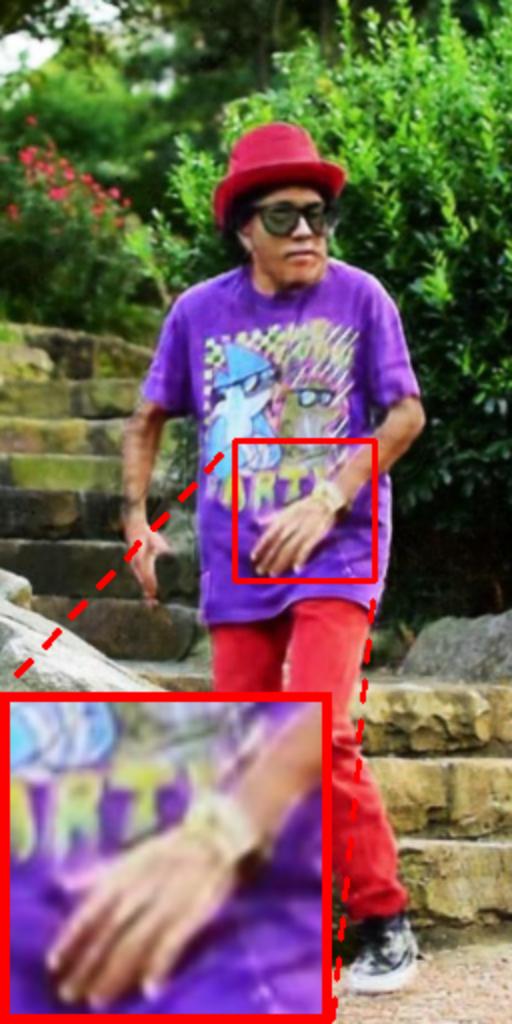}
    \end{tabular}
    \end{spacing}
    \caption{\sytneww{Comparisons of using Laplacian feature with alternative 3D constraints.  The first row visualizes different 3D constraints, the second row exhibits the direct transfer results of MT-Net, and the third row shows the detail-enhanced results of DE-Net. \sytminor{Note the inconsistent colors between wrists and hands caused by the discontinuous UV representation.}}}
    \label{fig:laplacian_feature}
\end{figure}

\sytneww{Note that Liu \etal~\cite{lwb2019} also use the projection of a 3D model to guide the generation of GAN. However, compared with the correspondence map colored by UV coordinates, our Laplacian features provide a more intrinsic and continuous representation for 3D models, which is more suitable for the human image generation task. }
\sytneww{Some other approaches, e.g., SMPLpix~\cite{Prokudin2021SMPLpixNA}, utilize the 3D geometry information by conditioning the depth map of human mesh vertices, which lacks structure information and is not suitable for the generative model.
The results in Table~\ref{table:compare_pose_label} and Fig.~\ref{fig:laplacian_feature} show that our approach leads to better scores and results.}

\begin{figure*}[t]
\centering
    \includegraphics[width=0.9\linewidth]{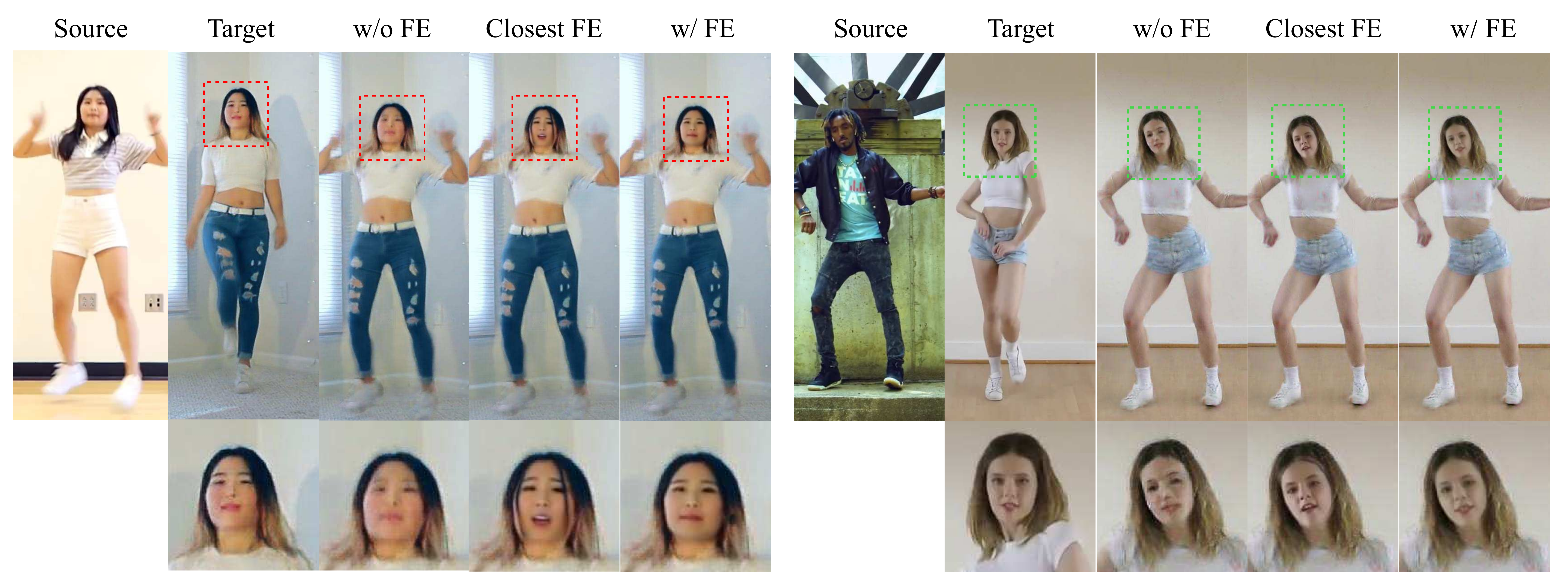}
    \vspace{-3mm}
    \caption{\sytneww{Effect of the facial enhancement step. It can be seen that our approach can help generate more faithful character faces. In contrast the ``Closest FE''
    baseline fails to keep the size or angle of faces, leading to fusion artifact (left) or incorrect poses (right).}}
    \vspace{-3mm}
    \label{fig:face_enhancement}
\end{figure*}

\sytneww{To demonstrate the effect of post-processing facial enhancement, we illustrate a pair of results with and without facial enhancement, denoted as \textbf{w/ FE} and \textbf{w/o FE}, respectively. Moreover, we also evaluate a baseline method, denoted as \textbf{Closest FE}, which is based on direct retrieval (i.e., enhancing the facial image by searching for the closest facial embedding in the latent space), instead of our retrieval-and-interpolation approach. 
The comparisons are illustrated in Fig.~\ref{fig:face_enhancement}.}

\subsubsection{User Evaluation}
\label{sec:user evaluation}
We also conduct\hbmajor{ed} a user study to measure the human perceptual quality for cross-subject transfer results. In our experiments, we compared videos generated by vid2vid, the method of Chan \etal, Liquid Warping GAN, and our method. Specifically, we showed to volunteers a series of videos by each of the methods at the resolution of $1024 \times 512$. \hbmajor{We invited 50 participants for this study.} 
\hbmajor{Without any time limit}, 
each of them 
\hbmajor{was} 
asked to select among results from the four approaches: 1) the clearest result with rich details; 2) the  temporally most stable result; and 3) the overall best result.
As shown in Table~\ref{table3}, our method \hbmajor{leads to more realistic results}, 
with richer details and with better temporal stability 
in comparison with other methods.

\begin{table}[h]
\vspace{-3mm}
\caption{User study results. We report the percentage of \hbmajor{the} participants' choice as the best result among the results of 4 methods in three different aspects, respectively.}
\vspace{-4mm}
\begin{center}
 \resizebox{\linewidth}{!}{
\begin{tabular}{l|l|llll}
\hline
\multicolumn{2}{c|}{\multirow{2}{*}{Quality}} & \multicolumn{4}{c}{Method} \\ \cline{3-6} 
\multicolumn{2}{c|}{} & vid2vid & chan \etal & LW-GAN & ours \\ \hline
\multicolumn{2}{l|}{\begin{tabular}[c]{@{}l@{}}Detail and clarity\end{tabular}} & 22.2\% & 16.7\% & 7.07\% & \textbf{54.0\%}  \\ \hline
\multicolumn{2}{l|}{\begin{tabular}[c]{@{}l@{}}Temporal stability\end{tabular}} & 24.7\% & 15.1\% & 9.60\% & \textbf{50.5}\%  \\ \hline
\multicolumn{2}{l|}{\begin{tabular}[c]{@{}l@{}}Overall feeling\end{tabular}} & 26.3\% & 17.7\% & 8.08\% & \textbf{48.0}\%  \\ \hline
\end{tabular}}
\end{center}
\vspace{-4mm}
\label{table3}
\end{table}

\begin{table}[h]
\vspace{-3mm}
\caption{Ablation study for pose labels. Our 3D condition achieves higher IS and lower FID scores,  indicating 
better results.}
\vspace{-6mm}
\begin{center}
 \resizebox{0.6\linewidth}{!}{
\begin{tabular}{l|l|l}
\hline
Metric   &   Pose labels in~\cite{lwb2019}  & Ours     \\ \hline
\sytneww{IS-ReID} $\uparrow$      &   \sytneww{3.612}                          & \sytneww{\textbf{4.015}}    \\ \hline
FID \hbmajor{$\downarrow$}      &   58.42                          & \textbf{51.26}    \\
\hline
\end{tabular}}
\end{center}
\vspace{-3mm}
\label{table:compare_pose_label}
\end{table}

\vspace{-6mm}
\SYT{
\subsubsection{Effect of Number of Source Frames on Results}
The previous motion transfer methods such as \cite{chan18:every_dance_now} and \cite{wang2018vid2vid} only use target frames at the training stage, and the quality of their generative models is \hbmajor{thus} not directly related to the number of source frames. While in our method, source and target frames are both involved in training. Therefore, it is meaningful to explore the influence of source frame number\hbmajor{s} on the \hbmajor{quality of} generated results. We carry out experiments on all subjects and record the average evaluation of generated image quality with respect to 
the ratio of source frame number to target when the number of target frames is fixed, as 
shown in Fig.~\ref{fig:loss}. We choose \hbmajor{the} inception score as the metric. It can be seen that the loss has converged when the ratio is around 0.5, which serves as a guidance for the data preparation.
}

\vspace{-3mm}
\subsection{Qualitative Results}
We visualize our generated results in Figure~\ref{fig:sample}. It can be seen that our method successfully drives the motions of different targets with structural integrity and rich details, particularly in the face and hands.
We also demonstrate that our method outperforms the existing methods in Figures~\ref{fig:contrast} and~\ref{fig:contrast_more}.

 As illustrated in the first row of Figure \ref{fig:contrast}, our method can maintain the structural integrity of human parts, and avoid the mixture of arms and hands, \hbmajor{which is often one of the main artifacts with the other methods.} 
 At the same time, our method can also characterize the details of the generated results more accurately, such as facial expression, \hbmajor{as} shown in the second row of Figure \ref{fig:contrast}. 
 Figure \ref{fig:scale} demonstrates the effectiveness of alignment transformation for DE-Net: it effectively aligns \hbmajor{the} source to the target, and thus improves the clarity and avoids artifacts in generated results.
 Figure \ref{fig:ablation} shows the advantage of using 3D constraints and DE-Net. 

\subsection{Limitations and \sytminor{Ethical} Discussions}
\begin{figure}[t]
\begin{center}
\includegraphics[scale=0.22]{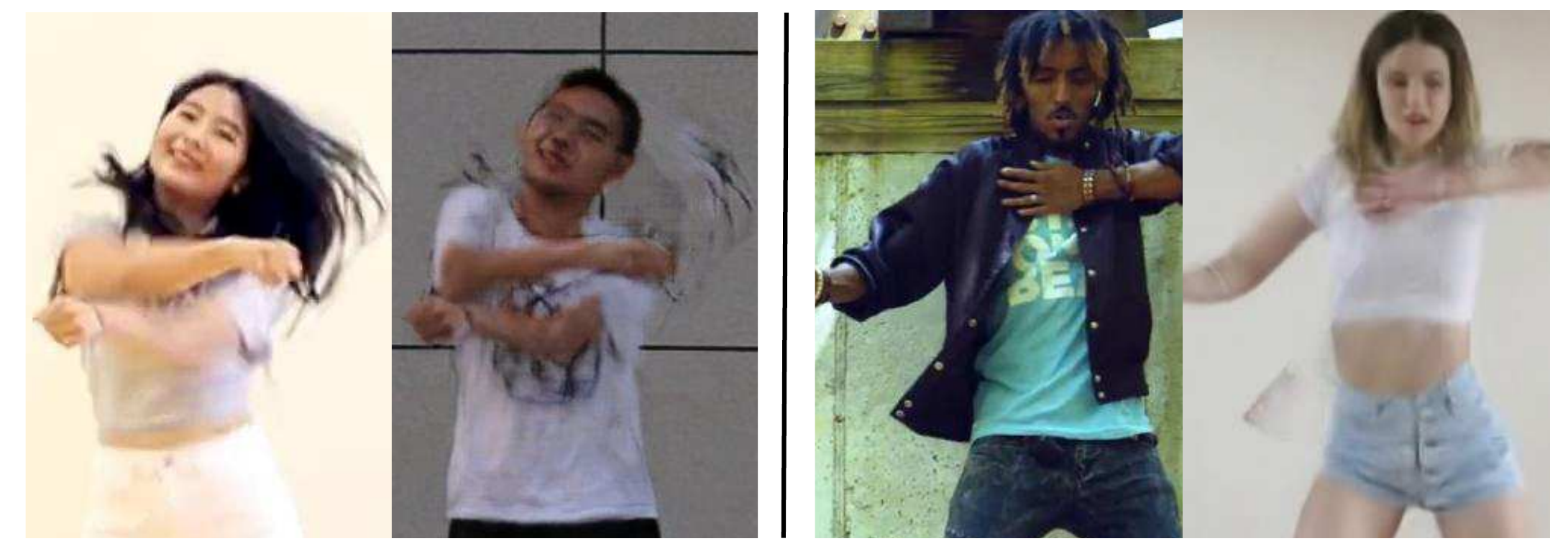}
\end{center}
\vspace{-3mm}
\caption{Failure cases. For each case, the source image is shown on the left and our transfer result on the right.}
\vspace{-6mm}
\label{fig:failure case}
\end{figure}

Although our model is able to synthesize motion transferred images with high authenticity and details, there are still several limitations. 
\sytminor{First, both the MT-Net and DE-Net are person-specific generation model\hbminor{s}. Therefore, \hbminor{their generalization ability is limited and they have} 
to be trained from scratch for any new source-target character pair.} 
\sytminor{Second, the abnormal movement of source characters might cause large changes in the human body shape, e.g., perturbations in hair or clothing, \hbminor{thus making}
the DE-Net fail to eliminate these suddenly emerging human parts.}
We show some failure cases with visual artifacts in Figure \ref{fig:failure case}. In the left example, our model fails to eliminate the long hair of the source character in the result, while in the right, some undesired part of clothes appears in the generated images because of the loosely dressed source subject. 

\sytminor{The intentions of our work are undoubtedly towards a comprehensive and positive perspective. However, potential harms might still be brought about if our technique is maliciously used. To reduce the impact of misuse (specifically, deep-fakes), previous works~\cite{yu2021survey} have proposed ways for detecting fake videos. We reasonably expect that these works could discriminate the generated videos \hbminor{by our method from} 
real ones and address concerns about the visual misinformation.}


 
\section{Conclusion}
We have proposed a new approach to human motion transfer. It employs the 3D body shape and pose constraints as a condition to regularize the generative adversarial learning framework, and the new condition
is more expressive and complete than 2D. We also design an enhancement mechanism to enhance the detailed characteristics of synthesized results using detailed information from  real source frames. Extensive experiments show that our method outperforms existing methods both qualitatively and quantitatively.

\section*{Acknowledgment}
This work was supported by the Beijing Municipal Natural Science Foundation for Distinguished Young Scholars (No. JQ21013), the National Key R\&D Program of China (No.2020AAA0104500), the National Natural Science Foundation of China (No. 62061136007 and No. 61872440), Royal Society Newton Advanced Fellowship (No. NAF\verb|\|R2\verb|\|192151), the Centre for Applied Computing and Interactive Media (ACIM) of School of Creative Media, City University of Hong Kong and the Youth Innovation Promotion Association CAS.

{\small
    \bibliographystyle{IEEEtran}
    \bibliography{IEEEabrv,egbib}
}

\end{document}